\documentclass[]{fairmeta}
\usepackage{xspace}
\newcommand{\ourmodel}{{eSEN-omol}\xspace}

\usepackage[version=4]{mhchem}

\usepackage{caption}
\usepackage{booktabs}
\usepackage{longtable}
\title{Quantum-accurate atomistic modeling of enzyme catalysis using a machine learned potential}

\author[1,*,\dagger]{Meng Gao}
\author[2,*]{Armin Shayesteh Zadeh}
\author[3]{Aniruddha Seal}
\author[2]{Siva Dasetty}
\author[2]{Siddarth K. Achar}
\author[1]{Misko Dzamba}
\author[1]{Benjamin K. Miller}
\author[1]{Leif D. Jacobson}
\author[1]{C. Lawrence Zitnick}
\author[1]{Brandon M. Wood}
\author[1,\dagger]{Zachary W. Ulissi}
\author[1,\dagger]{Daniel S. Levine}
\author[2,3,\dagger]{Andrew L. Ferguson}

\affiliation[1]{FAIR at Meta, San Francisco, California, United States}
\affiliation[2]{Pritzker School of Molecular Engineering, University of Chicago, 5640 South Ellis Avenue, Chicago, Illinois 60637, United States}
\affiliation[3]{Department of Chemistry, University of Chicago, 5735 South Ellis Avenue, Chicago, Illinois 60637, United States}

\contribution[*]{These authors contributed equally and are listed alphabetically}
\contribution[\dagger]{Co-corresponding authors}

\abstract{Electronic rearrangements associated with bond forming/breaking in catalytic enzymes require quantum mechanical (QM) treatment beyond classical molecular mechanics (MM). Hybrid QM/MM methods enable tractable simulations but require system-specific setup and are sensitive to the QM region choice and treatment of the QM/MM interface. We demonstrate quantum-accurate treatment of all-atom, complete enzymes in explicit solvent comprising up to 54k atoms and 1 microsecond of total simulation time using the machine-learned interatomic potential (MLIP) \ourmodel. We reproduce experimental barrier trends for Claisen rearrangement in chorismate mutase, resolve critical intermediate states in PETase catalyzed polymer depolymerization, and distinguish mechanistic alternatives for metal-activated phosphoryl transfer in nucleoside diphosphate kinase. We realize 1000$\times$ speedups relative to typical QM/MM calculations without system-specific tuning. These results establish MLIPs as a practical route to QM-accurate simulations of enzyme catalysis.}

\begin{document}
\maketitle

\section{Introduction}

Enzymes are molecular machines that catalyze the chemical reactions underlying life and play a central role in metabolism, signaling, and biotechnology\cite{hunter_protein_1995,bornscheuer_engineering_2012,robinson_enzymes_2015}. Scalable mechanistic simulations of enzyme catalysis remain a fundamental challenge because of the requirement for both quantum mechanical (QM) accuracy and the ability to simulate large systems over extended timescales, typically nanoseconds to microseconds \cite{Warshel2014Nobel}. Classical molecular mechanics (MM) force fields rely on a fixed bonding topology and thus cannot model changes in bonding. Reactive force fields permit such changes, but typically require bespoke parameterizations for each system, can miss subtle electronic effects, and tend to reproduce reference reaction energetics only qualitatively \cite{trnka_automated_2018}. 

This fundamental trade-off between QM accuracy and biomolecular scale has motivated a long history of approximate modeling \cite{lu_continuous_2025,marrink_two_2023,senn_qmmm_2009,kamerlin_empirical_2011, moerman_systematic_2021}. Traditional strategies to reduce complexity include truncated cluster models of the active site \cite{hutter2002mechanism}, which can incur errors from neglecting a large portion of the system, and hybrid QM/MM approaches, which embed a reactive QM region within a larger classical environment \cite{Warshel2014Nobel,senn_qmmm_2009}. 

However, such calculations necessitate system-specific knowledge and expert judgment regarding definition of the QM and MM regions, selection of QM functional and MM force fields, and treatment of the QM/MM boundary. This often requires extensive parameter tuning and can risk overfitting to experimental data \cite{senn_qmmm_2009,kulik2018qm/mm,Demapan2022,Cui2021}. The quality of QM/MM predictions is particularly sensitive to choice of the QM region: too small leads to unreliable results, too large becomes computationally prohibitive \cite{rivas2026efficientqmmm, Cui2021, kulik_how_2016}. Semi-empirical methods \cite{xtb1_grimme_2017, dftb3_2011} can reduce computational cost, typically at the expense of accuracy \cite{lonsdale_practical_2012,arantes_benchmark_2025}, although new methods are reshaping this trade-off \cite{gxtb_2025}. When solvent plays a role in the reaction, the manually chosen number of QM water molecules adds another parameter that can strongly influence results \cite{wilkins2023accurate,li2025accurate}. QM/MM still remains computationally expensive, with simulations typically limited to tens of picoseconds, \cite{li2025accurate,garciameseguer2023insights} making it challenging to observe barrier crossing events.

Machine-learned interatomic potentials (MLIPs) offer an appealing alternative. MLIPs are trained on QM datasets to replicate the performance of QM simulations with orders of magnitude lower computational cost. Large biomolecules represent challenging systems for MLIPs because of their chemically heterogeneous nature, complex many-body effects, and the requirement of stable long-time molecular dynamics (MD) integration. Recent specialized frameworks have shown great promise. Wang \textit{et al.}\ introduced AI$^2$BMD, which achieves \textit{ab initio} protein energies and forces using a fragmentation strategy and treats solvent using a polarizable model \cite{wang_ab_2024}. The GEMS and SO3LR frameworks are complementary routes to \textit{ab initio} accuracy for biomolecular simulations that have been demonstrated on protein folding and solvation dynamics in small systems \cite{kabylda_molecular_2025,unke_biomolecular_2024}. ML/MM approaches, in which an MLIP is used for the QM region, have thus been used to alleviate the simulation costs of QM/MM \cite{wang2026machine,ohmura2025mlmm,sun2026cmmutasemlmm}, but the challenges associated with ML region specification, MM force field choice, ML/MM boundary treatment, and solvent modeling all remain.  

\begin{figure*}[ht!]
    \centering
    \includegraphics[width=1.0\textwidth]{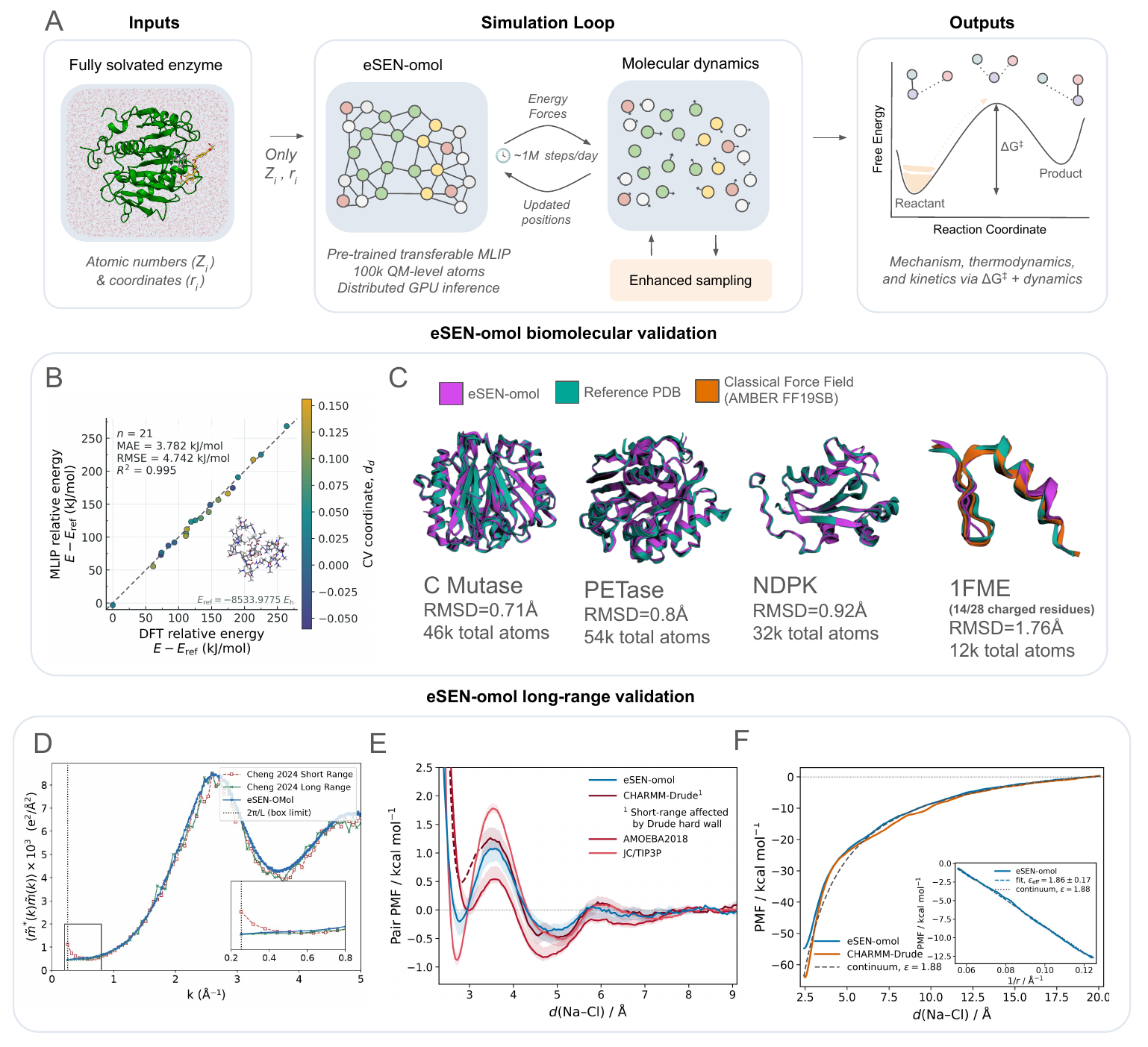}
    \caption{
    \textbf{Overview and validation of \ourmodel for enzyme catalysis simulations.} \textbf{(A)} Overview of our \ourmodel based enzyme workflow. Only atom coordinates $r_i$ and atomic numbers $Z_i$ are required as input; no bonding topology or QM-region/force-field/interface definition steps are necessary. \textbf{(B)} Parity plot showing that \ourmodel single point energies match DFT energies well for chorismate mutase clusters ($\sim$350 atoms) along the reaction coordinate. \textbf{(C)} Simulations of the three large proteins considered in this work and 1FME, a highly-charged peptide, are stable under \ourmodel simulations and remain close to their folded states (C$_\alpha$ RMSD < 2 \r{A}) over 2 ns of simulation with explicit solvent. \textbf{(D)} Dipole density correlation function following Ref.\ \cite{cheng2024latentewaldsummationmachine} shows that \ourmodel preserves dipole correlations in the long range limit and compares favorably to the ``long-range'' models reported therein. \textbf{(E)} Potential of mean force (PMF) of a single \ce{NaCl} ion pair in water (0.026 M) probed by umbrella sampling on the Na--Cl distance. The PMF correlations are largely damped by the high dielectric constant of water around $r_{cut}$ = 6 \r{A}, though features beyond that are resolvable. \textbf{(F)} Analogous PMF for an \ce{NaCl} ion pair in hexane (0.011 M), a low dielectric solvent. \ourmodel reproduces the predictions of the classical CHARMM force field and continuum predictions out to 20 \r{A} $>>$ $r_{cut}=6~\text{\AA}$. Fitting $1/r$ to the long range tail of the \ourmodel PMF accurately predicts the hexane dielectric constant as $\epsilon$ = 1.87 $\pm$ 0.17 (inset). See Appendix \ref{sec:suitability} for more details on \textbf{B-E}.
    }
    \label{fig:overview}
\end{figure*}

Recent reports indicate that models trained on the Open Molecules 2025 dataset \cite{levine_open_2025} match DFT performance in numerous chemical and biological applications \cite{husistein2026new, Kumar2026, SawyerWagen2025, ohmura2025mlmm}. We show here that \ourmodel\footnote{While UMA and \ourmodel perform similarly, we focus on \ourmodel as its lower cost to retrain allows for the ready creation of variants and ablations for testing hypotheses.}, the base model of UMA \cite{wood_uma_2025},
can readily scale and be effectively applied to full-chain enzyme catalysis in explicit solvent, treating all atoms at the same level of theory and removing the need for region and boundary selection (Figure \ref{fig:overview}A). \ourmodel is a message-passing MLIP trained on OMol25 at the $\omega$B97M-V/def2-TZVPD level of theory \cite{levine_open_2025}. We leverage established enhanced sampling protocols for rare-event simulation to promote transition state crossing and accurately estimate barrier heights in enzymatic reactions \cite{henin_enhanced_sampling_2022}. Our integrated distributed inference framework is able to run fully solvated enzyme simulations comprising 32-54k atoms on 8 GPUs. Additional GPUs accelerate simulations and with up to 32 GPUs simulation speeds of up to millions of steps per day on these systems can be realized (Appendix \ref{sec:distributed}) and total simulation times of $\sim$1.0 $\mu$s are aggregated in this work.

We demonstrate \ourmodel's suitability for biological simulations by reproducing sampled cluster DFT single-point energies with high-fidelity, correctly predicting long-range electrostatic behavior in scenarios relevant to proteins in solvent environments, and keeping proteins correctly folded over many nanoseconds of simulation. Figure \ref{fig:overview}B-E shows a collection of these validations with supporting details provided in Appendix \ref{sec:suitability}. We then study a diverse set of enzymatic systems with a range of different reactive mechanisms to recover free energy surfaces and reaction rate estimates with a single \ourmodel model, illustrating the accuracy\footnote{In this work, we highlight that our method achieves comparable accuracy to many past QM/MM results but we focus on the generalization capabilities of our model rather than the absolute accuracy that can be achieved by fitting to system specific parameters} and transferability of our method. Our results establish a route to QM-accurate simulations of full-chain, explicit solvent enzyme catalysis without system-specific parameterization or tuning (Figure \ref{fig:overview}A).

\section{Results and Discussion}

\subsection{Claisen Rearrangement in Chorismate Mutase} \label{subsec:CM}
Chorismate mutase (CM) is a textbook model for enzyme catalysis \cite{ray_kinetic_2024,claeyssens2011analysis}. It catalyzes the Claisen rearrangement of chorismate to prephenate, a key step in the biosynthesis of tyrosine and phenylalanine. This reaction also proceeds in aqueous solution via the same concerted pericyclic mechanism. Here, we study the \textit{Bacillus subtilis} CM enzyme (PDB: 3ZO8 \cite{pdb3zo8}).  Experimentally, the enzyme accelerates the reaction by $10^6$ to $10^7$-fold relative to aqueous solution \cite{kast1996chorismate}, with the rate enhancement attributed to preferential stabilization of the chair-like transition state through electrostatic interactions involving an arginine residue in the active site \cite{sogo1984stereochemistry,claeyssens2011analysis,burschowsky2014electrostatic}. 

\begin{figure*}[ht!]
    \centering
    \includegraphics[width=\textwidth]{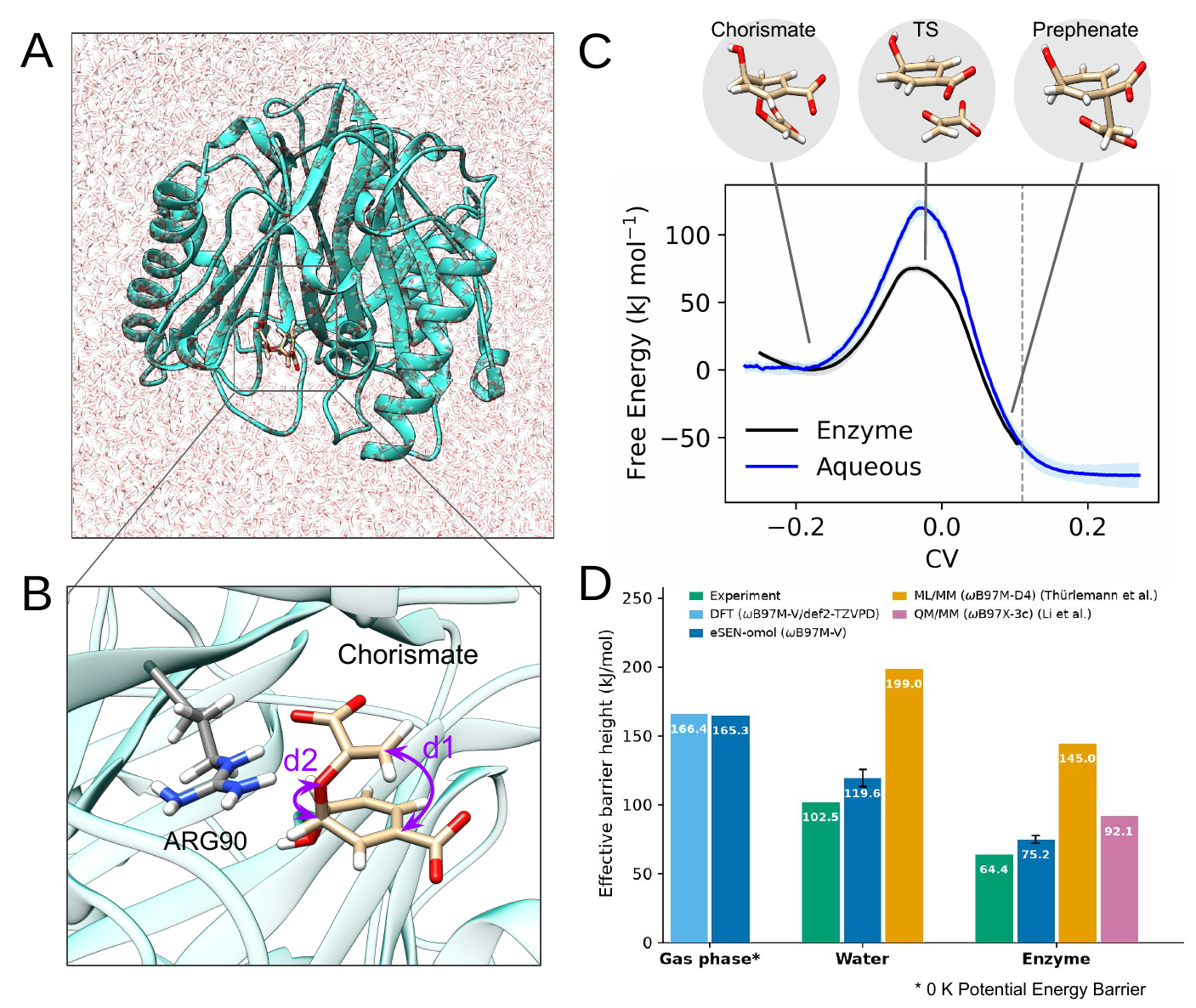}
    \caption{\textbf{Claisen rearrangement catalyzed by chorismate mutase.} \textbf{(A)} Structure of the full-chain chorismate mutase enzyme PDB: 3ZO8 \cite{pdb3zo8} in explicit water solvent (46k total atoms, 40k water atoms). \textbf{(B)} Close-up view of the chorismate mutase binding-pocket illustrating the Claisen rearrangement reaction where $d_1$ is the \ce{C-C} forming bond and $d_2$ is the \ce{C-O} breaking bond. ARG90 is one of the critical residues that forms H-bonds with the substrate. \textbf{(C)} Free energy surfaces for the Claisen rearrangement in water and enzyme binding pocket using \ourmodel with OPES. The enzyme-catalyzed free energy barrier calculated by \ourmodel is $\Delta G^\ddagger_\text{enzyme}$ = (75.21 $\pm$ 2.72)~$\mathrm{kJ~mol^{-1}}$ compared to $\Delta G^\ddagger_\text{water}$ = (119.81 $\pm$ 6.24)~$\mathrm{kJ~mol^{-1}}$ in water. \textbf{(D)} Comparison of simulated and experimental kinetics. Gas-phase barrier heights are the 0~K electronic energy differences between optimized reactant and transition state structures. For aqueous and enzyme-catalyzed systems, activation free energies were predicted from OPES calculations and verified by computing OPESf rate constants. Comparison is made only to previous studies using similar range-separated hybrid DFT functionals: the ML/MM values from Thürlemann \textit{et al.}\ \cite{thurlemann2026amp} with an ML model trained on $\omega$B97M-D4/ma-def2-TZVPP reference data, and the estimated QM/MM value with $\omega$B97X-3c reported by Li et al. 2025 (without CC refinement)\cite{li2025accurate} (More details in Appendix \ref{sec:c_mutase_structure_li_chan}). Experimental barrier data are taken from Kast \textit{et al.}\ \cite{kast1996chorismate}.}
    \label{fig:chorismate}
\end{figure*}

We first examined the intrinsic reactivity in the gas phase using QM calculations at the $\omega$B97M-V/def2-TZVPD level of theory. The QM potential energy profile agrees closely with \ourmodel predictions, with respective barrier heights of 169.8~$\mathrm{kJ~mol^{-1}}$ and 169.0~$\mathrm{kJ~mol^{-1}}$ (\Cref{fig:cm_dft}), providing an initial validation of the pre-trained MLIP. We then introduced explicit water solvent (\Cref{fig:chorismate}A) and conducted $\sim$46k-atom \ourmodel simulations at 300 K and 1 bar with On-the-fly Probability Enhanced Sampling (OPES) \cite{invernizzi_rethinking_2020,invernizzi_unified_2020} along a previously reported collective variable (CV) to drive the Claisen rearrangement (\Cref{fig:chorismate}B, Appendix \ref{sec:chorismate_appendix}) \cite{li2025accurate, ray_kinetic_2024}. The resulting free-energy surface (FES) in water gives a barrier of $\Delta G^\ddagger_\text{water}$ = (119.81 $\pm$ 6.24)~$\mathrm{kJ~mol^{-1}}$ (Figure \ref{fig:chorismate}) in good agreement with experiment (102.55~$\mathrm{kJ~mol^{-1}}$ \cite{andrews1973transition}) and previous computational studies (102.09 $\pm$ 2.51~$\mathrm{kJ~mol^{-1}}$ \cite{wilkins2023accurate}) (\Cref{fig:chorismate}D). Finally, we conducted OPES simulations of the reaction within the full-chain, explicitly-solvated enzyme environment and find that the pre-trained \ourmodel predicts a reaction barrier of $\Delta G^\ddagger_\text{enzyme}$ = (75.21 $\pm$ 2.72)~$\mathrm{kJ~mol^{-1}}$, in reasonable agreement with the experimentally-reported value of 64.43~$\mathrm{kJ~mol^{-1}}$ \cite{kast1996chorismate} (\Cref{fig:chorismate}C,D), and corresponding to a $\sim$45~$\mathrm{kJ~mol^{-1}}$ reduction compared to the barrier in water. The $\sim$10~$\mathrm{kJ~mol^{-1}}$ discrepancy with respect to experiment is comparable to the errors in the DFT training energies underpinning \ourmodel. Specifically, we found that the training level of theory $\omega$B97M-V/def2-TZVPD overestimates the CM gas phase barrier height by a similar $\sim$10~$\mathrm{kJ~mol^{-1}}$ compared to DLPNO-CCSD(T)/def2-QZVPPD. Others have seen similar errors for $\omega$B97M-V barrier heights \cite{liang_gold-standard_2025}. Further, the relative barrier reduction $\Delta\Delta G^\ddagger = \Delta G^\ddagger_\text{water} - \Delta G^\ddagger_\text{enzyme}$, which can mitigate systematic errors, differs by only $\sim$6.5~$\mathrm{kJ~mol^{-1}}$.

In addition, we computed the rate constants using OPES flooding (OPESf) \cite{ray2022rare}. In water, we obtained a rate constant of $k_\text{water}$ = ($9.01 \pm 4.53) \times 10^{-9}$ ~$\mathrm{s^{-1}}$ corresponding to an effective activation free energy of $E_a^\text{water}$ = (119.69 $\pm$ 1.28)~$\mathrm{kJ~mol^{-1}}$ (\Cref{fig:cm_opesf}). In the enzyme, the calculated rate constant increases to $k_\text{enzyme}$ = (0.26 $\pm$ 0.12)~$\mathrm{s^{-1}}$ implying an effective activation free energy of $E_a^\text{enzyme}$ = (76.77 $\pm$ 1.16)~$\mathrm{kJ~mol^{-1}}$. The calculated kinetic activation energies are in excellent agreement with the free energy barriers within $\sim$1.5~$\mathrm{kJ~mol^{-1}}$, providing strong support for the consistency of the independent thermodynamic and kinetic barrier height estimations.  The predicted $k_\text{enzyme}$/$k_\text{water}$ = $3.2 \times 10^{7}$ acceleration is in good agreement with the experimentally reported accelerations of $10^6-10^7$-fold \cite{kast1996chorismate}. These results (\Cref{fig:chorismate}D) constitute a marked improvement on replicating experimental barriers over previous ML/MM \cite{thurlemann2026amp} and match the best QM/MM \cite{li2025accurate} study that uses a density functional fine-tuned to (high level) electronic-structure theory, while improving on the QM/MM result with an untuned functional. While QM/MM methods in general can already achieve high accuracy~\cite{wilkins2023accurate, lichan2025pnas}, we envision \ourmodel's combination of speed, strong generalization without system specific tuning, and partitioning free setup will accelerate the field of enzymology and more broadly quantum-accurate simulations.

\subsection{Ester-Bond Hydrolysis in PETase}

\begin{figure*}
    \centering
    \includegraphics[width=1.0\textwidth]{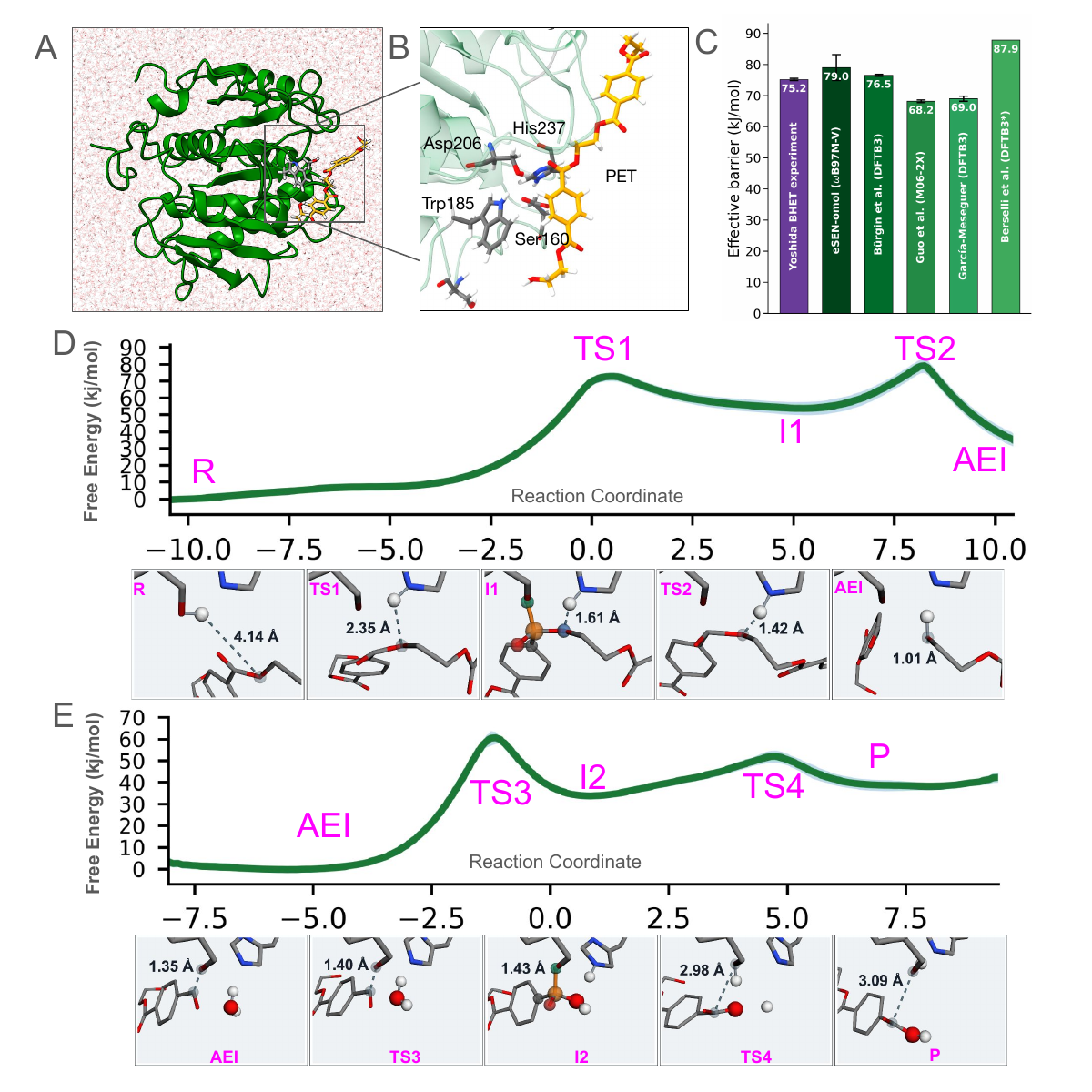}
    \caption{\textbf{Acylation and deacylation reactions during hydrolysis of a PET dimer substrate by \textit{Ideonella sakaiensis} PETase.} \textbf{(A)} Structure of the full-chain PETase structure from Burgin \textit{et al.}\ \cite{burgin2024} in explicit water solvent (54k total atoms, 50k water atoms). \textbf{(B)} Close-up view of the PETase catalytic residues and PET dimer. \textbf{(C)} Comparison of apparent barrier predictions to other QM/MM studies \cite{burgin2024, guo2025qmmm, garciameseguer2023insights, berselli2025molecular} and experimental measurements of BHET dimer hydrolysis \cite{yoshida2016bacterium,chen2018structural}. \textbf{(D-E)} The acylation and deacylation FES calculated by replica exchange umbrella sampling (REUS) (50 ns total simulation) using the reaction coordinates (RCs) (Appendix \ref{petase_rc}) from Burgin \textit{et al.}\ \cite{burgin2024}. We observe distinct, tetrahedral intermediate structures, stabilized by the enzyme, in both the acylation (I1) and deacylation (I2) reactions, consistent with the metastable intermediate states reported in Guo \textit{et al.} \cite{guo2025qmmm}, but not in Burgin \textit{et al.}\ \cite{burgin2024} despite our calculations using RCs from the latter. Our measured barriers place acylation at $(79.03 \pm 4.2)$ $\mathrm{kJ~mol^{-1}}$ and deacylation at $(60.73 \pm 3.0)$ $\mathrm{kJ~mol^{-1}}$, and support acylation as the rate limiting step.}
    \label{fig:petase}
\end{figure*}

The PETase enzyme from \textit{Ideonella sakaiensis} is a promising biocatalyst for depolymerization of poly(ethylene terephthalate) (PET) \cite{yoshida2016bacterium}. PETase employs the canonical catalytic SER-HIS-ASP triad for PET hydrolysis \cite{hedstrom2002serine, rauwerdink2015how}  (Figure~\ref{fig:petase}A,B). Structural and biochemical studies suggest that PET hydrolysis proceeds through sequential acylation and deacylation steps \cite{yoshida2016bacterium}. During acylation, the catalytic SER160 attacks the ester bond of PET to form an acyl-enzyme intermediate (AEI), which is subsequently hydrolyzed by a water molecule to regenerate the active enzyme while releasing the product during the deacylation step.
A number of QM/MM studies have probed the mechanism and free energy surface associated with this reaction \cite{burgin2024, berselli2025molecular, guo2025qmmm, knott2020characterization, jackering2024influence, boneta2021qmmm}, but differences in DFT settings \cite{jackering2024influence,berselli2025molecular}, selection of the QM region \cite{magalhaes2022, jackering2024influence}, length of the polymer chains \cite{guo2025qmmm}, and choice of CVs \cite{jackering2024influence,burgin2024} have led to debate on whether the acylation \cite{berselli2025molecular,garciameseguer2023insights} or deacylation \cite{burgin2024,guo2001substrate} step is rate limiting and whether each reaction proceeds in a single step \cite{burgin2024} or through a two-step tetrahedral intermediate \cite{guo2025qmmm}.

Here, we utilize the solvated PETase structure from Burgin \textit{et al.} (PDB: 6EQE), as well as the corresponding reaction coordinates (RCs) \cite{burgin2024, AustinPETasePDB}, and PET is represented by a bis(2-hydroxyethyl) terephthalate (BHET) dimer. We conducted $\sim$54k-atom replica exchange umbrella sampling (REUS) \cite{sugita2000multidimensional} to sample the acylation and deacylation pathways (Appendix \ref{sec:petase_reus}). Transition state region geometries are consistent with the transition state structures reported by Burgin \textit{et al.} \cite{burgin2024}. 

As illustrated in Figure \ref{fig:petase}C, we predict an effective free energy barrier of $(79.03 \pm 4.2)$ $\mathrm{kJ~mol^{-1}}$, derived from the acylation step, to be rate limiting, in good agreement with prior QM/MM studies \cite{burgin2024, guo2025qmmm, garciameseguer2023insights, berselli2025molecular} and experimental measurements for the BHET dimer \cite{yoshida2016bacterium}.
We assume, as have others \cite{burgin2024, guo2025qmmm, garciameseguer2023insights, berselli2025molecular}, that the acylation and deacylation steps are separable, as diffusion of BHET away from the enzyme effectively renders the acylation step irreversible.
Our results are in good agreement with experimental $k_\text{cat}$ derived barrier height of ($75.19 \pm 0.4$) $\mathrm{kJ~mol^{-1}}$ \cite{yoshida2016bacterium,chen2018structural} and that predicted by Burgin \textit{et al.}\ of 76.5 $\mathrm{kJ~mol^{-1}}$ \cite{burgin2024}. Compared to Burgin \textit{et al.}, however, our FES exposes some key mechanistic differences. 
First, our calculations predict acylation, rather than deacylation, to be rate-limiting, as was found by other studies \cite{garciameseguer2023insights, berselli2025molecular}. The highest free energy along the acylation reaction at transition state TS2 is $(79.03 \pm 4.2)$ $\mathrm{kJ~mol^{-1}}$ (Figure \ref{fig:petase}D, Table \ref{tab:petase-reus-features}), whereas the highest barrier along the deacylation reaction at transition state TS3 is slightly lower at $(60.72 \pm 3.0)$ $\mathrm{kJ~mol^{-1}}$ (Figure \ref{fig:petase}E, Table \ref{tab:petase-reus-features}).
Second, despite using the identical CV and similar umbrella sampling methods, we predict both acylation and deacylation to proceed by a two-step mechanism with well-defined intermediates as opposed to a single concerted step. The acylation intermediate I1 lies $\sim$19 $\mathrm{kJ~mol^{-1}}$ lower in free energy than TS1 and $\sim$25 $\mathrm{kJ~mol^{-1}}$ lower than TS2, splitting the nucleophilic addition of the SER160 side chain and the elimination of the BHET leaving group. The deacylation intermediate I2 lies $\sim$27 $\mathrm{kJ~mol^{-1}}$ lower in free energy than TS3 and $\sim$19 $\mathrm{kJ~mol^{-1}}$ lower than TS4, splitting the attack of catalytic water and the elimination of SER160 to regenerate the enzyme. Both intermediates show a distinct tetrahedral geometry centered on the PET ester carbonyl carbon, consistent with the simulations of Guo \textit{et al.} \cite{guo2025qmmm}, which employed a higher level of DFT theory (M06-2X/6-31G) than Burgin \textit{et al.} (DFTB3) \cite{burgin2024} who did not find this behavior. This observation is consistent with previous work showing that PBE and DFTB3 tend to underestimate reaction barriers and produce overly smooth free-energy surfaces, whereas methods incorporating exact exchange can better localize charge along bond-breaking and bond-forming coordinates and resolve finer features \cite{jackering2024influence, guo2025qmmm}. Taken together, our results suggest that \ourmodel, which is trained at the $\omega$B97M-V/def2-TZVPD level of theory, can accurately predict barrier heights and reaction mechanisms at $\sim$1000$\times$ speedups (Appendix \ref{sec:distributed}) relative to QM/MM methods with a high level of DFT theory. 

\subsection{Enzyme-catalyzed phosphoryl transfer in nucleoside diphosphate kinase}

NDPK is a metal-activated phosphotransferase that balances cellular pools of nucleoside triphosphates for signaling, metabolism, and mitochondrial function \cite{xu_alf3_1997,morera1994adenosine}. It operates through a ping--pong mechanism in which the active-site HIS122 captures the $\gamma$-phosphoryl group of a nucleoside triphosphate (NTP) such as ATP, and donates it to an incoming nucleoside diphosphate (NDP) \cite{de_la_rosa_nm23nucleoside_1995,schneider_mechanism_2001}. 
We model the NDPK dephosphorylation half-reaction in quantum-accurate simulations with full-chain proteins, explicit solvent, and a metal ion using \ourmodel. We initialized our \ourmodel simulations of NDPK from the \textit{Dictyostelium discoideum} crystal structure (PDB: 1KDN \cite{xu_alf3_1997}).

\begin{figure*}[h!t]
    \centering
    \includegraphics[width=1.0\textwidth]{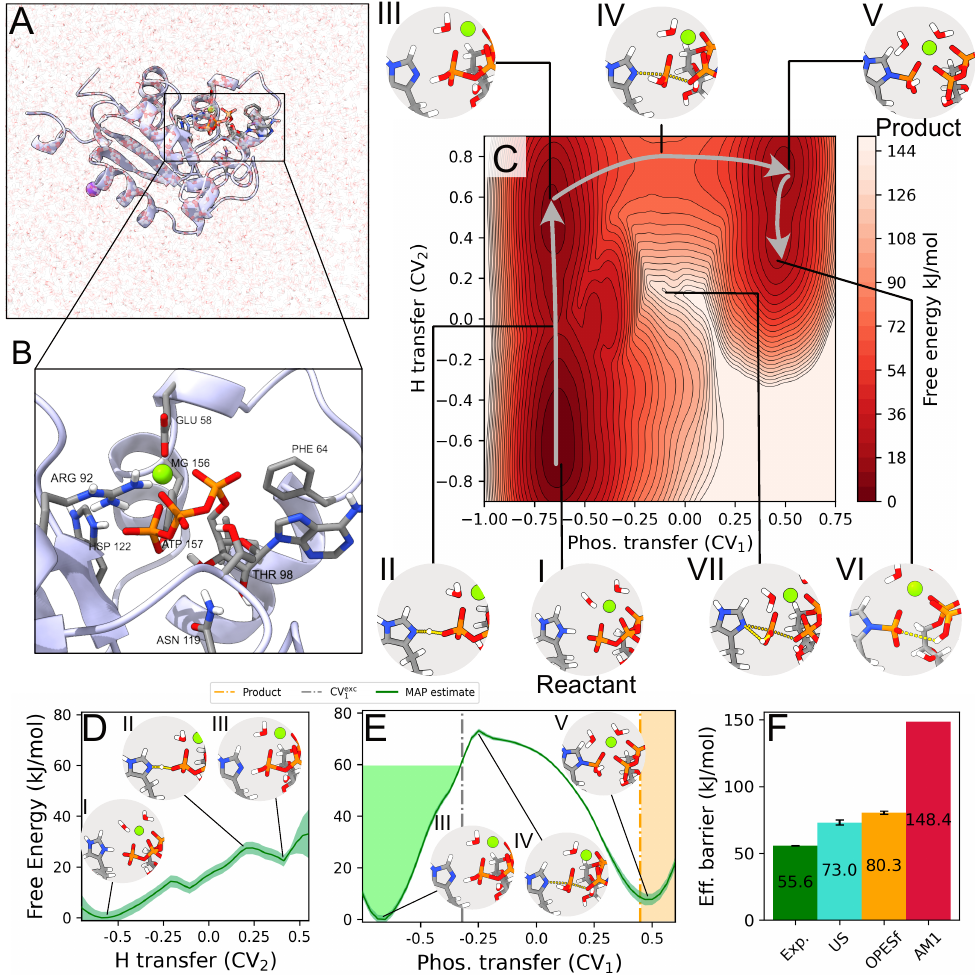}
    \caption{
      \textbf{NDPK phosphoryl transfer dephosphorylation half-reaction.} 
      \textbf{(A)} Representative structure of NDPK in explicit water solvent (32k total atoms, 30k water atoms).
      \textbf{(B)} Close-up view of the relaxed binding pocket showing ATP hydrogen-bond contacts to TYR56, ARG92, THR98, and ASN119 and octahedral hexacoordination of the Mg$^{2+}$ ion with the three phosphoryl groups of ATP, two water molecules, and GLU58. 
      \textbf{(C)} FES from 2D WTMetaD along the phosphoryl-transfer coordinate CV$_1$ and proton-transfer coordinate CV$_2$. The low-free energy pathway for the dephosphorylation half-reaction is predicted to proceed via a two-step mechanism comprising a proton transfer from the doubly-protonated HIS122 to a $\gamma$-phosphoryl oxygen (I$\to$II$\to$III) followed by transfer of the protonated $\gamma$-phosphoryl group to N$\delta$ of HIS122 through an S$_\mathrm{N}$2-like transition state. Upon completion of the phosphoryl transfer reaction, the hydrogen on the phosphorylated HIS is free to transfer back and forth between the negatively charged oxygen atoms of the $\beta$-phosphoryl group on ADP and the phosphorylated HIS (V$\to$VI).
      \textbf{(D,E)} FES from 1D US for proton transfer and phosphoryl transfer with corresponding free energy barriers of $\Delta G^\ddagger_1$ = (27.49 $\pm$ 1.05)~$\mathrm{kJ~mol^{-1}}$ and $\Delta G^\ddagger_2$ = (73.05 $\pm$ 1.86)~$\mathrm{kJ~mol^{-1}}$.
      \textbf{(F)} Comparison of estimated barriers from experiment \cite{gonin1999catalytic}, US, OPESf, and the AM1 cluster-model benchmark \cite{hutter2002mechanism}.
      }
    \label{fig:ndpk}
\end{figure*}

We calculated the FES of the phosphoryl transfer reaction using well-tempered metadynamics (WTMetaD) \cite{wtmetad} driven in two collective variables: CV$_1$ tracks nucleophilic attack of HIS122 N$\delta$ on the $\gamma$-phosphorus, and CV$_2$ tracks proton transfer from N$\delta$ of HIS122 to the $\gamma$-phosphoryl oxygen (Appendix \ref{sec:NDPK_appendix}). The FES suggests that the low-free energy pathway corresponds to a two-step mechanism rather than the concerted proton/phosphoryl transfer previously hypothesized \cite{hutter2002mechanism} (\Cref{fig:ndpk}C). In the first step, the proton on the doubly-protonated HIS122 hops to a $\gamma$-phosphoryl oxygen (I$\to$II$\to$III). The protonated $\gamma$-phosphoryl group then transfers through an S$_\mathrm{N}$2-like transition state to N$\delta$ of HIS122, completing the dephosphorylation half-reaction (III$\to$IV$\to$V). Once the phosphorylation reaction is completed, the proton previously transferred from HIS122 to $\gamma$-phosphoryl is free to hop back and forth between the $\beta$-phosphoryl on ADP and the phosphoryl group on HIS122 (V$\to$VI) with a barrier of just $\sim$7.5~$\mathrm{kJ~mol^{-1}}$. The transition state (VII) of the concerted proton/phosphoryl transfer mechanism (I$\to$VII$\to$V) stands $\sim$100~$\mathrm{kJ~mol^{-1}}$ above the reactant, which is substantially higher than the corresponding value of $\sim$70~$\mathrm{kJ~mol^{-1}}$ for the transition state (IV) of the rate-limiting second step of the two-step pathway (III$\to$IV$\to$V). We note that, as is the case for all low-dimensional FES projections, barrier heights depend on the choice of CVs and care should be taken not to overinterpret reaction paths and activation energies. Nevertheless, the CVs here were chosen from physical understanding of the mechanism, resolve both the two-step and concerted pathways within the same FES, and identify the two-step pathway as lower in free energy. As we show below, we observe good agreement of the predicted barrier height for the putative rate limiting step with an OPESf rate calculation. 

As an independent verification of the free energy barriers calculated from 2D WTMetaD, we conducted 1D umbrella sampling (US) \cite{torrie1977nonphysical} of the two elementary steps in the proposed two-step pathway. US calculations in CV$_2$ between states I and III expose proton hopping as a largely uphill transition through the transition state II with an elementary free energy barrier of $\Delta G^\ddagger_1$ = (27.49 $\pm$ 1.05)~$\mathrm{kJ~mol^{-1}}$ (\Cref{fig:ndpk}D). An analogous US calculation in CV$_1$ between states III and V resolves phosphoryl transfer to be a single-barrier two-state mechanism with an elementary free energy barrier of $\Delta G^\ddagger_2$ = (73.05 $\pm$ 1.86)~$\mathrm{kJ~mol^{-1}}$ (\Cref{fig:ndpk}E). The barriers computed by 1D US are in agreement with those computed by marginalization over the reaction channels in the 2D WTMetaD FES (Figure \ref{fig:ndpk_margin}). The overall free energy change for the dephosphorylation half-reaction between states I and V is approximately $\Delta G$ = 15~$\mathrm{kJ~mol^{-1}}$.

We applied OPESf to the rate limiting phosphoryl transfer step (\Cref{fig:ndpk}E, \Cref{sec:rate_eqn}) \cite{ray2022rare,seal_computing_2025} and obtained a predicted rate constant of $k_\text{OPESf}$ = (0.043 $\pm$ 0.019)~$\mathrm{s^{-1}}$ with a corresponding apparent effective activation free energy of $E_a^\text{OPESf}$ = (81.32 $\pm$ 1.13)~$\mathrm{kJ~mol^{-1}}$ (\Cref{fig:ks_npdk}). The effective activation energy agrees with the free energy barrier computed by US to within $\sim$10~$\mathrm{kJ~mol^{-1}}$. In \Cref{fig:ndpk}F we compare our US (73~$\mathrm{kJ~mol^{-1}}$) and OPESf (80.3~$\mathrm{kJ~mol^{-1}}$) barriers to the value implied by the experimental $k_\mathrm{cat}$ (55.6~$\mathrm{kJ~mol^{-1}}$ \cite{gonin1999catalytic}). Because $k_\mathrm{cat}$ reports on the full cycle, which is diffusion-limited, this value provides an upper bound on the chemical barrier. The US estimate shows the best agreement with experiment, exceeding it by $\sim$17~$\mathrm{kJ~mol^{-1}}$, representing a roughly five-fold error reduction from the previous semi-empirical AM1 cluster-model (148.4~$\mathrm{kJ~mol^{-1}}$ \cite{hutter2002mechanism}).

\section{Conclusion}

The ability to perform quantum-mechanically accurate simulations of full-chain, explicit-solvent enzyme catalysis without system specific setup or tuning has the potential to transform mechanistic understanding and rational design of enzyme catalysis. In this work, we demonstrate that pre-trained MLIPs provide a viable pathway to this goal. \ourmodel, trained on OMol25, simutaenously offers QM-accurate calculations at the $\omega$B97M-V/def2-TZVPD level of theory and $\sim$1000$\times$ speedups over equivalent QM/MM methods (Appendix \ref{sec:distributed}), as well as treatment of full-chain enzymes in explicit solvent at a unified level of theory without any system-specific setup and tuning. We couple the \ourmodel simulations to off-the-shelf rare event enhanced sampling techniques -- REUS \cite{sugita2000multidimensional}, WTMetaD \cite{wtmetad}, OPES \cite{invernizzi_unified_2020}, and OPESf \cite{ray2022rare} -- to efficiently estimate free energy surfaces, barriers, and rates. In an application to chorismate mutase, we accurately predicted effective barriers and rate accelerations consistent with experimental measurements. In PETase, our rate-limiting barriers are in good agreement with experiment and prior QM/MM calculations and we resolve previously reported metastable intermediates. In the NDPK dephosphorylation half-reaction, we predicted a two-step reaction mechanism mediated by interstitial solvent molecules and a \ce{Mg^2+} cation in which proton transfer is followed by a rate-limiting phosphoryl transfer, yielding a reaction barrier with an approximately five-fold lower deviation from experiment than a previous semi-empirical QM study.

While transferable MLIPs have demonstrated great promise across broad areas of chemistry \cite{levine_open_2025} and we demonstrated here their promise in biomolecular simulations of enzymatic catalysis, these methods are still relatively new and have certain current limitations and open questions. MLIPs inherit any shortcomings that exist in the training data, including systematic errors associated with the level of DFT theory and statistical errors associated with the diversity of the training set. MLIPs are also generally untested on systems which may require careful treatment of significant charge transfer or multiple electronic or spin states. Unlike classical force fields, long-range electrostatics and induced polarization are implicitly encoded in a graph neural network and not directly accessible for modulation. Further, while MLIPs are much faster than QM calculations, they remain significantly slower than classical potentials, so enhanced sampling techniques with their attendant complexities are still required for MLIP-driven simulations of enzyme catalysis.

The applications in this work have demonstrated that MLIPs can open new avenues for scientific discovery in mechanistic understanding and computational design of enzymes going beyond the prediction of stable structures \cite{jumper_highly_2021,abramson_accurate_2024,krishna2024generalized,lin_evolutionary-scale_2023} and conformational ensembles \cite{lewis_scalable_2025,jing_alphafold_2024,jin2025p2dflow,Janson2025aa} to fully embrace the dynamical complexity of enzyme catalysis. In the NDPK application, for example, \ourmodel enabled the prediction of complex reaction mechanisms that would otherwise remain difficult to model or suffer from strong biases associated with system-specific setup. As these methods continue to mature, they promise to transform our ability not only to understand complex biomolecular processes, but also to systematically design systems with targeted functions. 

\section*{Acknowledgements}
We thank Juno Nam for fruitful conversations and critical feedback on an early draft of the manuscript. We additionally thank Bingqing Cheng for providing dipole correlation data. Calculations were completed in part with resources provided by the University of Chicago Research Computing Center. We gratefully acknowledge computing time on the University of Chicago high-performance GPU-based cyberinfrastructure supported by the National Science Foundation under Grant No.\ DMR-1828629. 

\textbf{Funding:} SKA acknowledges support from the Eric and Wendy Schmidt AI in Science Fellowship, a Schmidt Sciences program.

\textbf{Author contributions:} 

Conceptualization: ASZ, CLZ, DSL, ALF;

Data curation: ASZ, MG, AS, SD, SKA, MD, BMW, DSL;

Formal analysis: ASZ, MG, AS, SD, SKA;

Investigation: ASZ, MG, AS, SD, SKA, MD, BKM, CLZ, BMW, ZWU, DSL, ALF;

Methodology: ASZ, MG, AS, SD, SKA, ALF, BMW;

Project administration: MG, ZWU, DSL, ALF;

Resources: CLZ, ZWU;

Software: MG, MD, BMW;

Supervision: CLZ, ZWU, ALF;

Validation: ASZ, MG, AS, SD, SKA, BMW, DSL;

Visualization: ASZ, MG, AS, SD, SKA, BMW;

Writing – original draft: ASZ, MG, AS, SD, SKA, MD, BMW, DSL, ALF;

Writing – review \& editing: MG, BMW, DSL, ALF.

\textbf{Competing interests:} ALF\ is a co-founder and consultant of Evozyne, Inc.\ and a co-author of US Patent Applications 16/887,710 and 17/642,582, US Provisional Patent Applications 62/853,919, 62/900,420, 63/314,898, 63/479,378, 63/479,378, 63/521,617, 63/510,130, 63/669,836, and 63/987,554 and International Patent Applications PCT/US2020/035206, PCT/US2020/050466, PCT/US24/10805, PCT/US24/34369, and PCT/US25/35833.

\textbf{Data, code, and materials availability}: 
All data are available in the manuscript or the supplementary materials.
The eSEN models are available at \url{https://huggingface.co/facebook/OMol25}. Code for running models is available at \url{https://github.com/facebookresearch/fairchem}.

\section*{Supplementary Materials}
Materials and Methods \\
Supplementary Text, Figures, Tables, and References \\

\bibliographystyle{sciencemag}
\bibliography{paper}

\clearpage
\newpage

\beginappendix

\setcounter{figure}{0}
\renewcommand{\thefigure}{S\arabic{figure}}

\setcounter{table}{0}
\renewcommand{\thetable}{S\arabic{table}}

\setcounter{equation}{0}
\renewcommand{\theequation}{S\arabic{equation}}

\section{Materials and Methods}

\subsection{Molecular Dynamics Simulations Using \ourmodel}\label{sec:md_methods}

Molecular modeling calculations were conducted using the ASE suite \cite{ase-paper} with the \ourmodel MLIP through Meta's fairchem package \cite{metafairchem}. Enhanced sampling simulations were performed with \texttt{PLUMED 2.11.0} enhanced sampling libraries \cite{plumed}. 

\subsection{Distributed Inference}\label{sec:distributed}
In order to perform enhanced sampling/molecular dynamics of these large enzymatic systems, we need to be able to run simulations on nanosecond/day timescales with O(100k) atoms with a highly accurate model for molecules. There are many MLIPs that can run simulations on this number of atoms but they are either not accurate enough for this application, do not support parallelism \cite{orbv3}, rely on strictly local interactions \cite{allegro, allegrofm}, or are limited to inter-node only parallelism \cite{distmlp}. Towards this end, we designed our domain decomposed parallelism algorithms to scale \ourmodel to millions of steps/day (equivalent to ns/day for 1 fs timesteps) speeds for 100k atom systems with any simulator engine including ASE and LAMMPS. The code is freely available on GitHub at \url{https://github.com/facebookresearch/fairchem/}. Execution speeds with ASE are reported in Table \ref{tab:speed-benchmark-spatial}.

In this paper, we reference 1000x speedup relative to high level QM/MM methods (for example: M06-2x, $\omega$B97X-3c). This is an approximate argument based on what is frequently reported in QM/MM studies where methods with $\omega$B97X-3c report 3000-6000 steps per day\cite{li2025accurate} and M06-2x report total simulation times of approximately tens of ps\cite{guo2025qmmm} for example.

\begin{table}[ht!]
  \centering
  \caption{
    Execution speeds of \ourmodel using ASE under standard NVT
    dynamics (1 fs timestep, Langevin thermostat at
    300 K), measured with the \texttt{turbo}
    inference settings on \texttt{Nvidia h200 140GB GPUs} nodes using all-to-all
    graph parallelism with spatial graph partitioning. Values are inferences
    per day, equal to MD steps per day. OOM denotes out-of-memory.
    }
  \label{tab:speed-benchmark-spatial}
\begin{tabular}{l r r r r} \toprule & & \multicolumn{3}{c}{Throughput ($10^6$ steps/day)} \\ \cmidrule(lr){3-5} System & Atoms & 1 GPU & 8 GPU & 32 GPU \\ \midrule Water (1000 mol.) & 3{,}000 & 0.86 & 3.17 & 3.27 \\ NDPK & 32{,}339 & \textsc{oom} & 0.60 & 1.53 \\ Chorismate mutase & 46{,}323 & \textsc{oom} & 0.38 & 1.06 \\ PETase (acylation) & 53{,}816 & \textsc{oom} & 0.35 & 0.97 \\ \bottomrule \end{tabular}

  \vspace{2pt}
  {\footnotesize
   \begin{minipage}{0.92\linewidth}
    \textit{Note.} Single-GPU runs for the solvated enzymes exhaust device
    memory, as expected at these system sizes.
   \end{minipage}}
\end{table}

\subsection{Suitability of \ourmodel for Large Scale Biological Simulations}\label{sec:suitability}

\subsubsection{DFT Accuracy}

We demonstrate that \ourmodel matches DFT to within a few kJ/mol accuracy in a 25 \AA\ sampled QM region for configurations harvested from an enzyme catalyzed reactive transition from chorismate to prephenate by the chorismate mutase enzyme under a 1.42 ns, 300 K OPES metadynamics trajectory (Figure \ref{fig:overview}B, cf.\ Section \ref{subsec:CM}). Defining the collective variable $\Delta d=d(\mathrm{C16-C3})-d(\mathrm{O11-C1})$, where the two terms describe the forming of C-C and breaking of C-O bonds in the CHB substrate, fifty target values were uniformly spaced between $\Delta d$ = (-0.060)-0.156 nm. For each target, the saved trajectory frame having the nearest CV value was selected and each frame was permitted to be used only once. Frames were available every 500 MD steps (0.25 ps). The selected configurations spanned 17.0-1400.75 ps, and the maximum deviation from a target CV value was $(5.33\times10^{-4})$ nm.
A fixed-composition active-site cluster was constructed from each configuration. Every cluster contained the complete CHB substrate; protein residues (denoted by chain name and residue) B57–B60, B63, B73–B75, C7–C9, C78, C90, and C108; and one conserved water molecule (SPC 8334). The catalytic residues were selected following previous QM/MM descriptions of the inter-subunit active site \cite{claeyssens_high-accuracy_2006,ranaghan_conformational_2004}, with additional persistent contact and packing residues included subject to a 350-atom computational limit. Pro B58 and Gly C8 were retained to avoid introducing adjacent, overlapping peptide caps. The water molecule was the most persistent water near CHB; its oxygen-to-substrate-heavy-atom distance ranged from 2.55 to 5.00 Å among the selected configurations.
The retained protein regions formed seven peptide fragments. Each fragment was terminated with an acetyl group at its N terminus and an N-methylamide group at its C terminus. Cap heavy-atom positions were inherited from the instantaneous geometry of the neighboring backbone, while cap hydrogens were placed using idealized bond lengths and tetrahedral geometry. All other coordinates were transferred directly from the trajectory without geometry optimization. Each resulting cluster contained 343 atoms, had a formal charge of +1, and was treated as a singlet.
The atom composition and ordering were identical for all configurations. As validation, the CV was recomputed directly from each trajectory frame and agreed with the corresponding COLVAR value within $(2\times10^{-6})$~nm. Hydrogen connectivity was checked at every selected geometry, and no interatomic separation below 0.70~\AA\ was permitted.
These CV-stratified snapshots were randomly subsampled to choose 24 structures. Note that orthogonal environmental coordinates are uncontrolled, so absolute cluster energies should not be interpreted directly as a one-dimensional reaction profile.

\begin{figure}[ht!]
\centering
    \includegraphics[width=0.9\textwidth]{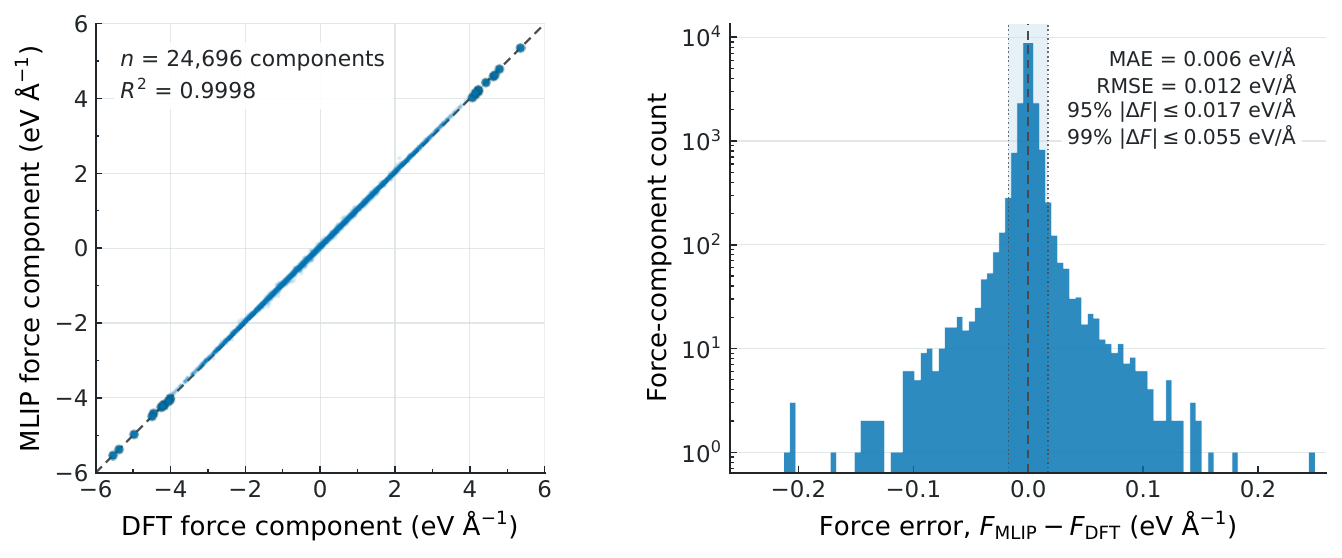}
    \caption{Force parity plot for eSEN-omol vs DFT ($\omega$B97M-V/def2-TZVPD) for chorismate mutase}
    \label{fig:force_parity}
\end{figure}

\subsubsection{Simulation Stability}

\begin{figure}[ht!]
\centering
    \includegraphics[width=0.6\textwidth]{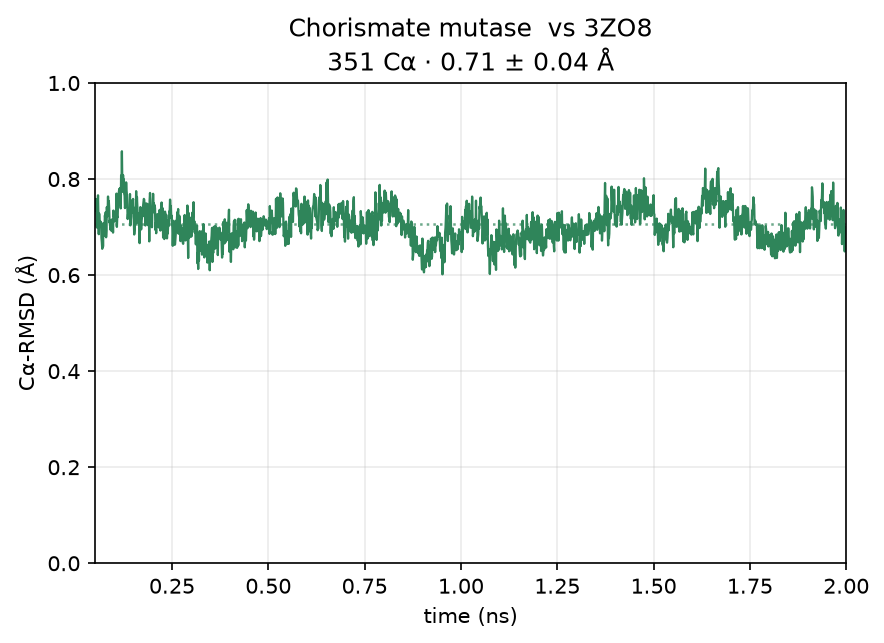}
    \caption{Trajectories of 2 ns NVT simulations with \ourmodel show chorismate mutase to remain stable and in agreement with the PDB: 3ZO8 \cite{pdb3zo8} structure to within a C$_\alpha$ RMSD of ($0.71 \pm 0.04$) \AA.}
    \label{fig:cmutase_rmsd}
\end{figure}

We also demonstrate that all-atom simulations of full-chain chorismate mutase, PETase, and NDPK proteins in explicit solvent and containing a significant number of charged residues, charged substrates, and ions using \ourmodel are stable and remain properly folded (C$_\alpha$ RMSD < 2 \r{A}) over multi-nanosecond time scales (Figure \ref{fig:overview}C, \ref{fig:cmutase_rmsd}, \ref{fig:petase_rmsd}, and \ref{fig:ndpk_rmsd}). While the time scales of these unbiased calculations are insufficiently long to observe meaningful conformational changes in large biomolecular systems, the stability of MLIPs over nanosecond time scales is not necessarily given or even assessed in many applications \cite{jin2024improving}. 

\subsubsection{Evidence for Correct Long Range Behavior}

\paragraph{Highly charged peptide.}
A 2~ns simulation of the highly charged 1FME peptide \cite{sarisky2001beta} comprising 14 charged residues (out of 28 total) and numerous long-range salt bridges produces a close match to experimental NMR structures (Figure \ref{fig:overview}C) and classical force field results (AMBER FF19SB). We solvated 1FME in water (12k atoms in total), minimized the energy, equilibrated 100 ps of NVT at 300K, 100 ps of NPT at 1~atm and then ran unbiased simulations for 2~ns. The C$_\alpha$ RMSD is averaged over the run.

\paragraph{Dipole density correlations.}
We show that our model produces the correct long range dipole correlations in water (Figure \ref{fig:overview}D). Replicating the calculations from Ref.\ \cite{cheng2024latentewaldsummationmachine}, we perform NVT simulations of a water box with 512 atoms over 2~ns and show the long range dipole correlation function $\langle \tilde{m}^\ast(k) \tilde{m}(k) \rangle$, where $\tilde{m}$ is the Fourier transform of the molecular dipole density, agrees with the long-range models with explicit Ewald summation out to the $2\pi/L$ k-space limit. 
Specifically, we compute the molecular dipole density of water,
\begin{equation}
\mathbf m(\mathbf r)=\sum_j \boldsymbol\mu_j \delta(\mathbf r-\mathbf r_j),
\end{equation}
where $\boldsymbol\mu_j$ is the dipole of water molecule $j$ and $\mathbf r_j$ is the position of the water oxygen atom.
For each allowed wavevector in reciprocal space, $\mathbf k=\frac{2\pi}{L}(n_x,n_y,n_z)$, we take the dipole moment projection,
\begin{equation}
    \mu_{j,L}=\boldsymbol\mu_j\cdot\hat{\mathbf k},
  \qquad
  \hat{\mathbf k}=\frac{\mathbf k}{|\mathbf k|},
\end{equation}
then we take the Fourier transform to compute the dipole density in k-space,
\begin{equation}
\widetilde m_L(\mathbf k) = \sum_j(\boldsymbol\mu_j\cdot\hat{\mathbf k}) e^{-i\mathbf k\cdot\mathbf r_j},
\end{equation}
where the reported value is averaged over trajectory frames. k-vectors of the same magnitude are averaged together. In order to draw quantitative comparisons with \cite{cheng2024latentewaldsummationmachine}, we use the same box size with 512 atoms and a length of 24.86 \AA\ which limits the smallest k computed to be 0.25 \AA$^{-1}$. We then overlaid Cheng's results on ours, shifting the spectra by a constant offset along the y-axis to align the first peak. 

\paragraph{PMF of ions in solution.}
We show that \ourmodel correctly captures the long-range behavior of ions in solution. The high dielectric constant of water ($\epsilon \approx 78$ at room temperature) means that charge interactions are screened rapidly over short distances and the potential of mean force (PMF) of a \ce{NaCl} ion pair closely agrees with force fields with explicit long range electrostatics terms (Figure \ref{fig:overview}E). In a low dielectric solvent like hexane, we show that \ourmodel captures electrostatic interactions in the \ce{NaCl} ion pair out to >20 \r{A} ionic separation (at which point the potential is effectively 0 kJ/mol) in excellent agreement with both the continuum limit and classical simulations using the CHARMM-Drude force field (Figure \ref{fig:overview}F).

The classical reference system contained one Na$^+$, one Cl$^{-}$, and 900 water molecules in a fixed 30 \AA\ cubic cell at 300 K. Three classical models were considered: Joung–Cheatham ions with rigid TIP3P water, polarizable AMOEBA2018 water and ions, and CHARMM-Drude ions with SWM4-NDP water. Umbrella sampling employed 16 windows spanning Na–Cl separations of 2.0–9.0 Å, with a force constant of 20 kJ mol$^{-1}$ \AA$^{-2}$. Following 100 ps of restrained equilibration, production lengths were 2 ns, 1 ns, and 0.5 ns per window for JC/TIP3P, AMOEBA2018, and CHARMM-Drude, respectively.

In low-dielectric \textit{n}-hexane, the reference system contained one NaCl pair and 691 hexane molecules in a fixed 53.25 \AA\ cubic cell at 300 K. The classical calculation used the polarizable CHARMM-Drude 2023 force field with matching ion and all-atom hexane parameters, PME electrostatics, a 12 \AA\ nonbonded cutoff, and a 10 \AA\ switching distance. Umbrella windows covered separations from 2.5 to 20 \AA, using progressively weaker restraints and wider spacing at larger separations; each window included 25 ps of restrained equilibration and 300 ps of production. The \ourmodel and CHARMM-Drude PMFs show similar distance dependence across the sampled range (Figure 2E). Moreover, fitting \ourmodel's PMF, to an $A/r + C$ functional form over 8–18 \AA\ gave, gave an effective dielectric constant of approximately 1.87, consistent with the experimental continuum value of 1.88 for hexane.

All PMFs were reconstructed using BayesWHAM \cite{bayeswham}, corrected for the radial Jacobian by adding $2k_\mathrm{B}T\ln r$, and referenced to zero over 8–9 \AA\ in water and 19–20 \AA\ in hexane. Uncertainties were estimated using time-block analyses or moving-block bootstrap reconstructions. The zero value near 20 \AA\ is therefore a chosen free-energy reference, rather than evidence that the physical ion–ion interaction has vanished completely. The short-range CHARMM-Drude water PMF below 3.25 \AA\ was treated as qualitative because of activity of the model’s Drude-particle hard wall.

In Appendices \ref{app:water} and \ref{app:polyala}, we provide additional assessments of the performance of \ourmodel in predicting the structure and dynamics of liquid water and the folding of a small peptide.

Taken together, these results provide support that \ourmodel is well-suited to protein-scale, condensed phase calculations in bulk water. Despite not having an explicit Coulombic energy term commonly found in classical force fields, it accurately captures long range electrostatic interactions over its effective receptive field of $r_{\textit{eff}}$ = 24 \AA\ that is sufficient to capture highly screened charge interactions in water solvent.  

\subsection{Rate Estimation with Eyring's Transition State Theory and OPES Flooding}\label{sec:rate_eqn}

To quantify reaction kinetics, we employed transition-state theory (TST) and OPES flooding (OPESf). Within Eyring's TST, the rate constant associated with an activation free-energy barrier $\Delta G^\ddagger$ is given by,
\begin{equation}
k_{\mathrm{TST}}=\kappa \frac{k_{\mathrm B}T}{h}\exp\left(-\frac{\Delta G^\ddagger}{RT}\right),
\end{equation}
where $\kappa$ is the transmission coefficient, here assumed to be $\kappa=1$, $k_{\mathrm B}$ is Boltzmann's constant, $h$ is Planck's constant, and $T$ is the temperature. The equilibrium free-energy barriers $\Delta G^\ddagger$ obtained from enhanced sampling calculations neglect dynamical recrossings of the transition-state dividing surface. Accordingly, we computed reaction rates using OPESf, which directly samples barrier-crossing events and thereby incorporates dynamical effects beyond the equilibrium TST estimate.

OPESf accelerates rare transitions by progressively flooding the reactant basin while leaving the transition-state region unbiased \cite{ray_rare_2022}. A history-dependent bias $V_b(s,t)$ is deposited along a collective variable $s$, while an excluded region prevents bias deposition beyond a system-specific threshold $s_{\mathrm{exc}}$. The maximum bias is controlled by the energy cutoff $\Delta E$. The parameters $\Delta E$ and $s_{\mathrm{exc}}$ were selected so as to reduce the effective barrier within the reactant basin while preserving unbiased dynamics in the transition-state region.

Provided that the transition-state region remains unaffected by the bias, the biased first-passage time $t_f$ can be rescaled to the corresponding unbiased transition time $t^*$ according to,
\begin{equation}
t^*=t_f\left\langle e^{\beta V_b(\mathbf{s})}\right\rangle_{U+V_b},
\label{eqn}
\end{equation}
where $\beta=1/k_{\mathrm B}T$, $U$ is the unbiased potential energy, and the ensemble average defines the acceleration factor associated with the flooding bias \cite{grubmuller_predicting_1995,voter_method_1997, tiwary_metadynamics_2013,ray_rare_2022,ray_kinetics_2023}. For a rare-event process, the distribution of unbiased transition times is expected to follow a Poisson process with cumulative distribution function,
\begin{equation}
P_{n\geq 1}(t)=1-\exp\left(-\frac{t}{\tau}\right),
\end{equation}
where $\tau$ is the characteristic transition time. The reaction rate was obtained as,
\begin{equation}
k_{\mathrm{OPESf}} = \frac{1}{\tau},
\end{equation}
by fitting the distribution of $t^*$ values obtained from an ensemble of independent OPESf trajectories. The consistency of the transition-time distribution with a Poisson process was assessed using a two-sample Kolmogorov--Smirnov test \cite{salvalaglio2014assessing,tiwary_how_2017,ray_kinetics_2023}. 

Uncertainties in the OPESf transition times and rate constants were estimated following the procedure described in Ref.\ \cite{seal_computing_2025}. The 95\% confidence interval $(\tau_l,\tau_u)$ for the characteristic transition time was obtained using Kaminsky's method,
\begin{equation}
\left(\frac{2\sum_{i=1}^{n} t_i}{\chi^2_{2n}(0.975)},;\frac{2\sum_{i=1}^{n} t_i}{\chi^2_{2n}(0.025)}\right),
\end{equation}
where $t_i$ is the rescaled transition time for the $i$th trajectory and $\chi^2_{2n}(\alpha)$ denotes the $\alpha$-quantile of the chi-squared distribution with $2n$ degrees of freedom. A symmetric uncertainty was reported as,
\begin{equation}
\Delta\tau=\frac{\tau_u-\tau_l}{2}.
\end{equation}
Since $k=1/\tau$, the corresponding confidence limits on the rate are,
\begin{equation}
k_l=\frac{1}{\tau_u},\qquad k_u=\frac{1}{\tau_l},
\end{equation}
with the reported symmetric uncertainty,
\begin{equation}
\Delta k=\frac{k_u-k_l}{2}.
\end{equation}

The OPESf rate constants were converted to effective activation  barriers by inverting the Eyring equation,
\begin{equation}
\Delta G^\ddagger_{\mathrm{eff}}=RT\ln\left(\frac{k_{\mathrm B}T}{h \cdot k_{\mathrm{OPESf}}}\right).
\end{equation}
The confidence limits on the effective barrier were obtained as,
\begin{equation}
\Delta G^\ddagger_{l}=RT\ln\left(\frac{ k_{\mathrm B}T}{h k_{u}}\right), \qquad \Delta G^\ddagger_{u}=RT\ln\left(\frac{k_{\mathrm B}T}{h k_{l}}\right).
\end{equation}
A symmetric uncertainty in the effective activation barrier was then reported as,
\begin{equation}
\Delta\left(\Delta G^\ddagger_{\mathrm{eff}}\right)=\frac{\Delta G^\ddagger_{u}-\Delta G^\ddagger_{l}}{2} = \frac{RT}{2}\ln\left(\frac{k_u}{k_l}\right)
\end{equation}
such that the OPESf-derived effective barrier is expressed as
$\Delta G^\ddagger_{\mathrm{eff}}\pm\Delta(\Delta G^\ddagger_{\mathrm{eff}})$.

The OPESf parameters employed for the different systems are summarized in Table~\ref{tab}.

\begin{table}[ht]
\centering
\caption{OPESf parameters and product-state definitions employed for the three systems. $\Delta E$ and \texttt{EXCLUDED\_REGION} are PLUMED parameters controlling the bias ceiling and the region in which bias deposition is excluded, respectively. The product region defines the boundary used to identify reactive trajectories.}
\label{tab}
\begin{tabular}{lcccc}
\toprule
System
& Collective variable
& $\Delta E$ (kJ mol$^{-1}$)
& Product region
& \texttt{EXCLUDED\_REGION} \\
\midrule

CM (Aqueous)
& $\mathrm{CV}$
& 120
& $\mathrm{CV} > 0.15~\mathrm{nm}$
& $\mathrm{CV} < -0.15~\mathrm{nm}$ \\

CM (Enzyme) &
$\mathrm{CV}$
& 60
& $\mathrm{CV} > 0.15~\mathrm{nm}$
& $\mathrm{CV} < -0.15~\mathrm{nm}$ \\

NDPK &
$\mathrm{CV}_1$
& 60
& $\mathrm{CV}_1 > 0.4~\mathrm{nm}$
& $\mathrm{CV}_1 < -0.32~\mathrm{nm}$ \\
\hline
\end{tabular}
\end{table}

\clearpage

\section{Chorismate Mutase}\label{sec:chorismate_appendix}

\subsection{Simulation Setup and Details}\label{methods:cm}

Initial structures for the full enzyme were obtained from Ray \textit{et al.}\ \cite{ray2024kinetic}, with the protein conformation based on PDB 3ZO8 \cite{pdb3zo8}. For the corresponding aqueous simulations, the system consisted only of the reactive chorismate (CHB) species, with the protein removed and the solute fully solvated in water. In all chorismate mutase simulations, we employed ASE \cite{ase-paper} together with \ourmodel \cite{wood_uma_2025}. Prior to the OPES simulations, each system was first energy minimized and equilibrated. For energy minimization, we used a force tolerance of 0.05 eV/\AA. We performed a two-stage equilibration, first in the NVT ensemble at $T=300$ K, followed by equilibration in the NPT ensemble at $T=300$ K and $P=1$ bar. We employed a Langevin thermostat during the equilibration and final scale simulations and an isotropic MTKNPT barostat during the NPT equilibration.

\subsection{Enhanced Sampling Calculations}
Following the reaction coordinate used previously for OPES flooding calculations of chorismate mutase \cite{li2025accurate}, the Claisen rearrangement of chorismate to prephenate was described using a single CV defined as $\text{CV}=d_1-d_2$, where $d_1$ is the distance associated with formation of the new C--C bond and $d_2$ is the distance associated with cleavage of the C--O bond (Figure~\ref{fig:chorismate}). Progression from reactant to product is captured by the concerted decrease in $d_1$ and increase in $d_2$. OPES simulations were performed at 300 K using a bias deposition pace of 500 steps and a bias barrier of 75 kJ mol$^{-1}$, with the adaptive kernel width updated every 1000 steps. Harmonic walls were applied along the reaction coordinate to restrict sampling to the region relevant to the forward reaction, with a lower wall at $\text{CV}=(-0.05)$ nm and an upper wall at $\text{CV}=0.27$ nm. Both walls used a force constant of $5.0\times10^{4}$ kJ mol$^{-1}$ nm$^{-2}$. The same OPES setup, but without the lower wall, was used for the corresponding aqueous-phase reaction, in which the reactive substrate was simulated in explicit solvent without the enzyme.

\begin{figure}[ht!]
    \centering
    \includegraphics[width=\linewidth]{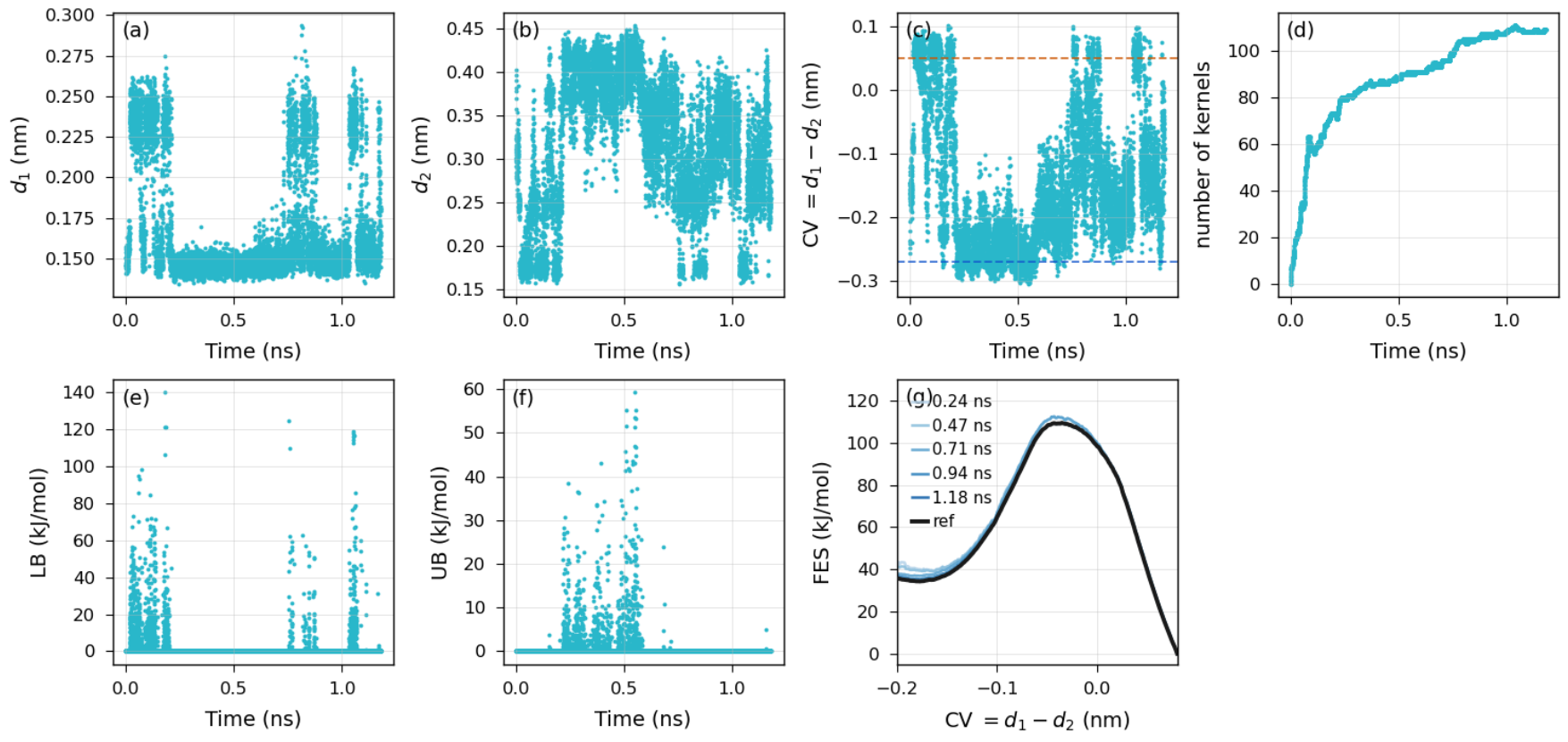}
    \caption{Convergence analysis of the OPES metadynamics simulation for chorismate mutase. Time evolution of (a)~$d_1$, (b)~$d_2$, (c)~$\mathrm{CV}=d_1-d_2$, (d)~number of deposited OPES kernels, (e)~lower harmonic wall bias (LB), (f)~upper harmonic wall bias (UB), and (g)~free-energy surfaces reweighted from cumulative trajectory fractions of 20\%--100\% (blue, light to dark) together with the full-data reference (black).}
    \label{fig:cm_conv}
\end{figure}

Convergence of the OPES simulation is assessed in
Figure~\ref{fig:cm_conv}. The distances $d_1$, $d_2$, and the CV undergo multiple spontaneous transitions between the reactant and product basins throughout the 1.18 ns trajectory (Figure~\ref{fig:cm_conv}a--c), satisfying the multiple-crossing criterion for sampling convergence~\cite{invernizzi_exploration_2022,invernizzi_unified_2020}. The number of deposited OPES kernels plateaus at ${\sim}111$ (Figure~\ref{fig:cm_conv}d), indicating that the adaptive bias has ceased to grow and the underlying free-energy landscape is well sampled. The lower and upper harmonic wall biases (LB and UB; Figure~\ref{fig:cm_conv}e--f) are activated only transiently, confirming that the OPES bias alone drives barrier crossing and the walls serve exclusively as boundary safeguards. Free-energy surfaces reweighted from five cumulative fractions of the trajectory (20\%--100\%) converge to a consistent barrier of ${\sim}109$\,kJ\,mol$^{-1}$, with all estimates within ${\sim}3$\,kJ\,mol$^{-1}$ of the full-data reference
(Figure~\ref{fig:cm_conv}g).

Going beyond equilibrium free energies, we next computed rate constants using OPES flooding (OPESf) \cite{ray_rare_2022}. The reweighted first-passage times of ensembles of 15 and 18 independent OPESf trajectories collected, respectively, for the aqueous and enzyme-catalyzed reactions follow Poisson statistics, from which we extracted estimates of the unbiased reaction rates (\Cref{sec:rate_eqn}, \Cref{fig:cm_opesf}) \cite{ray2022rare,seal_computing_2025}. In water, we computed a rate constant of $k_\text{water}$ = ($9.01 \pm 4.53) \times 10^{-9}$ ~$\mathrm{s^{-1}}$ corresponding to an effective activation free energy of $E_a^\text{water}$ = (119.69 $\pm$ 1.28)~$\mathrm{kJ~mol^{-1}}$ (\Cref{fig:cm_opesf}). In the enzyme, the calculated rate constant increases to $k_\text{enzyme}$ = (0.26 $\pm$ 0.12)~$\mathrm{s^{-1}}$ with a corresponding effective activation free energy of $E_a^\text{enzyme}$ = (76.77 $\pm$ 1.16)~$\mathrm{kJ~mol^{-1}}$. The calculated kinetic activation energies are in excellent agreement with the free energy barriers within $\sim$1.5~$\mathrm{kJ~mol^{-1}}$, providing strong support for the consistency of the independent thermodynamic and kinetic barrier height estimations.  Our kinetic calculations predict a $k_\text{enzyme}$/$k_\text{water}$ = $3.2 \times 10^{7}$ acceleration of the enzyme-catalyzed reaction relative to aqueous solution, in good agreement with the experimentally reported accelerations of $10^6$ to $10^7$-fold \cite{kast1996chorismate}.

\subsection{Structural Model Variation and Comparison to Li and Chan Results}\label{sec:c_mutase_structure_li_chan}
We note that when we attempted to dock chorismate into the PDB: 2CHT \cite{pdb2cht} crystal structure, we initially obtained an improperly docked system in which the reaction center is largely solvent exposed and the substrate is only stabilized by a single hydrogen bond (\Cref{fig:cm_hb}). Our \ourmodel OPESf calculations for this structure yield a similarly elevated free energy barrier of 108~$\mathrm{kJ~mol^{-1}}$ (\Cref{fig:cm_pdb_comp}) compared to Li and Chan's results on their 1HB structure\cite{li2025accurate}. In contrast, the higher-resolution PDB: 3ZO8 \cite{pdb3zo8} structure resulted in a more stably docked substrate, featuring a two-hydrogen-bonded substrate conformation and a substantially lower activation barrier, directly supporting the original observations from Li and Chan\cite{li2025accurate}. This result also highlights that while studies of reaction dynamics are now readily feasible with MLIPs, an accurate initial structural model is still a vital first step.

We also note that in Figure \ref{fig:chorismate}, we only compared to Li and Chan's values on $\omega$B97X-3c to keep the level of theory and functional comparable to \ourmodel. Li and Chan showed that refining $\omega$B97X-3c parameters on LNO-CCSD(T) on chorismate single points indeed improved their results. In Section \ref{sec:c_mutase_gas_phase}, we find that computing barriers with DLPNO-CCSD(T) also yields similar improvements, suggesting that future models should be trained on CCSD(T) data when they become available.

\subsection{\emph{Ab initio} validation of gas phase barrier height prediction}\label{sec:c_mutase_gas_phase}

\begin{figure}[ht!]
    \centering
    \includegraphics[width=0.5\linewidth]{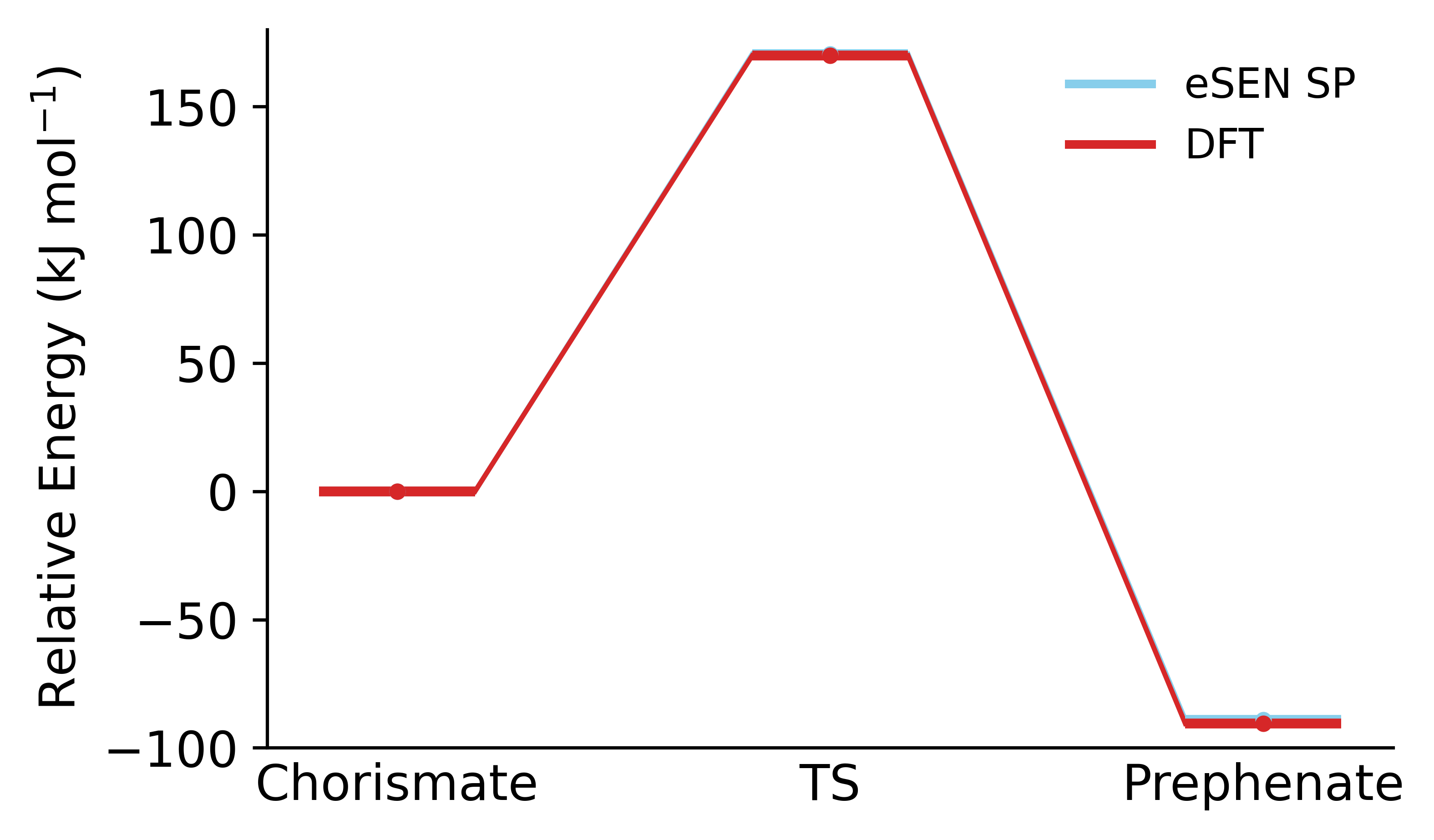}
    \caption{Gas-phase reaction profile for chorismate-to-prephenate conversion computed with \ourmodel and KS-DFT using $\omega$B97M-V/def2-TZVPD in ORCA on structures obtained from the growing string method.}
    \label{fig:cm_dft}
\end{figure}

\begin{figure}[ht!]
    \centering
    \includegraphics[width=0.5\linewidth]{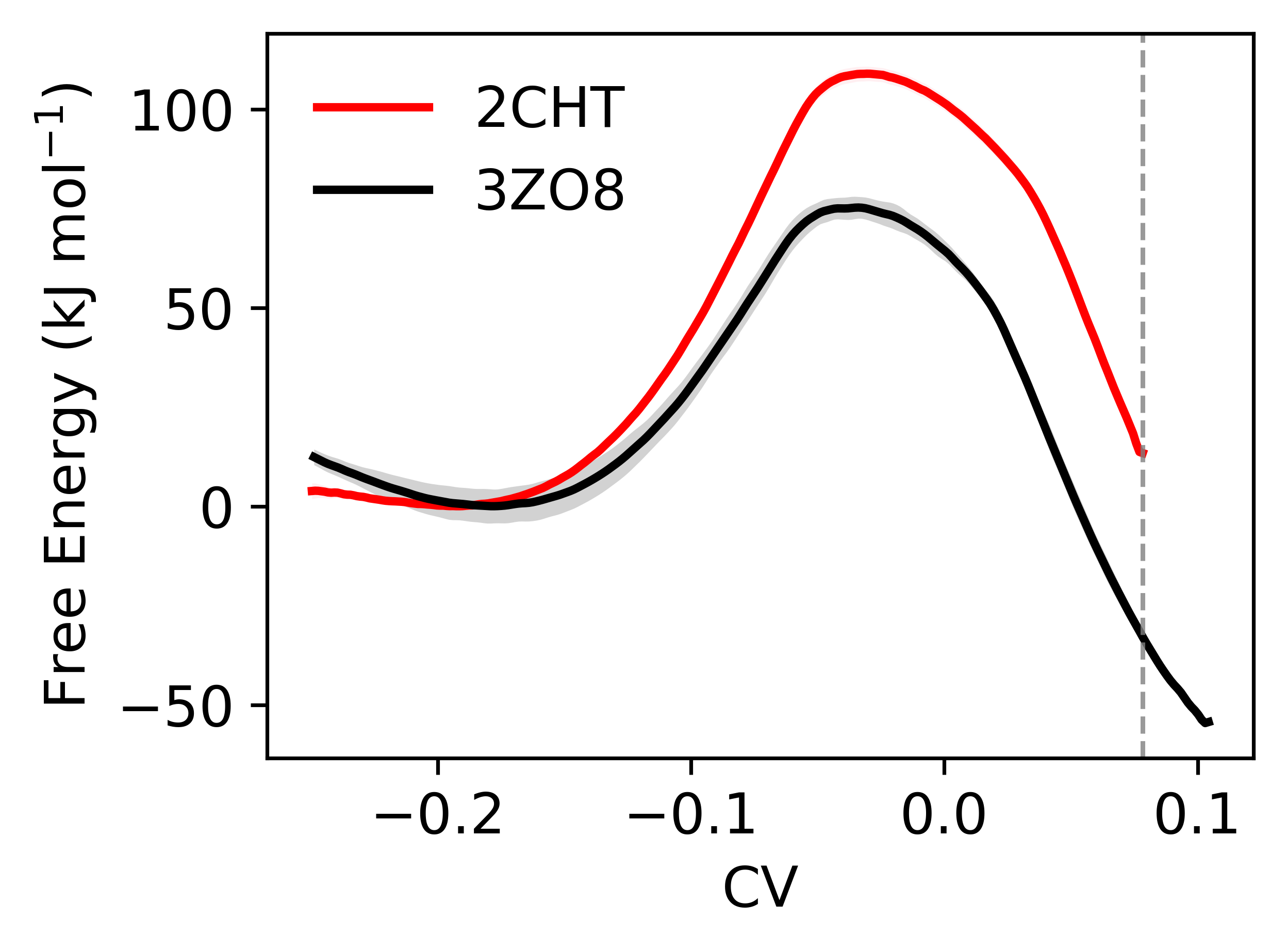}
    \caption{Free-energy surfaces for the reaction in two fully solvated enzymes constructed with PDB ID: 3ZO8 (reported in main text) and 2CHT, computed with OPES along the collective variable $\text{CV}=d_1-d_2$.}
    \label{fig:cm_pdb_comp}
\end{figure}

\begin{figure}[ht!]
    \centering
    \includegraphics[width=0.8\textwidth]{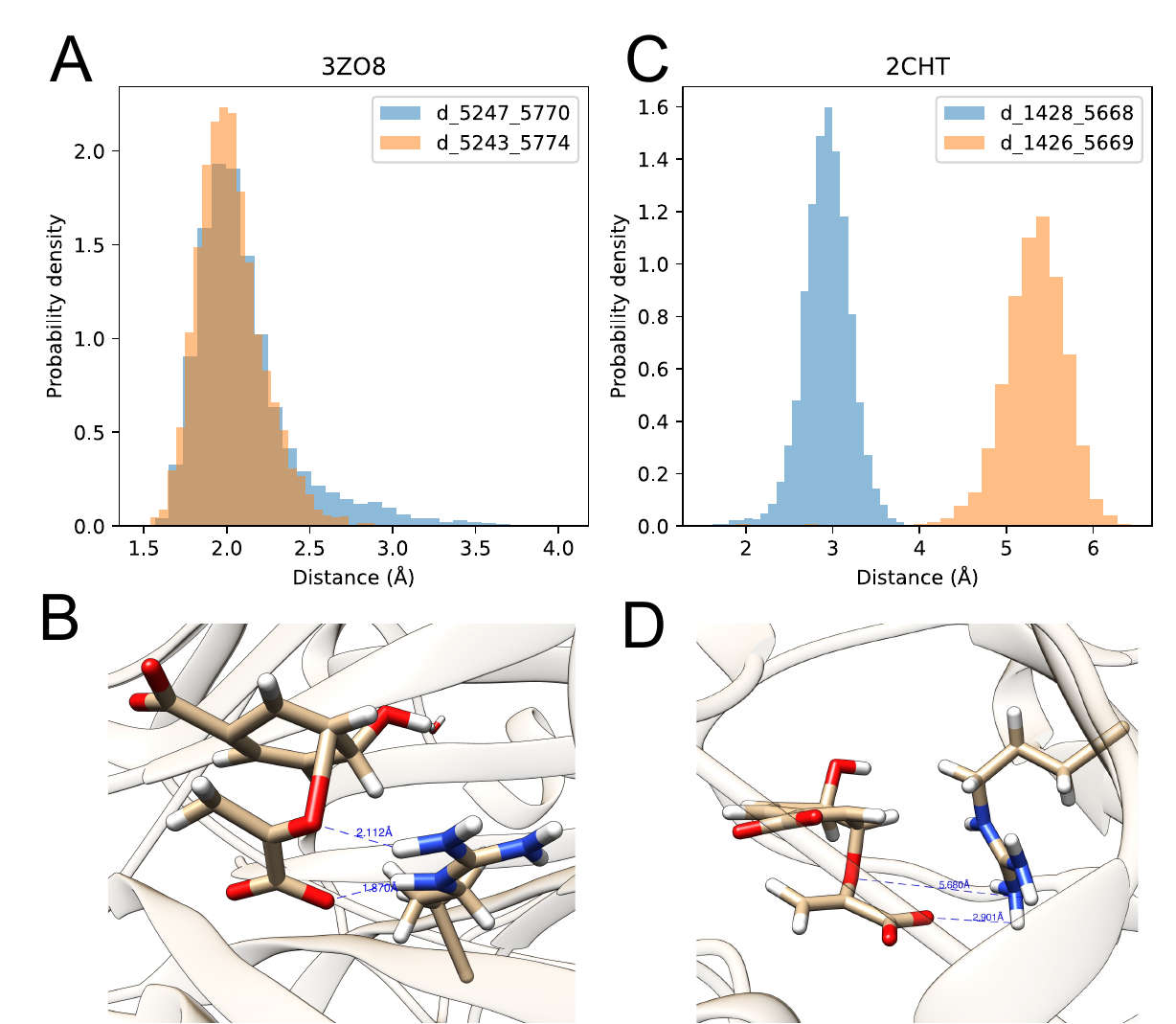}
    \caption{H-bonding distance distributions between O atoms in the CHB residue and H atoms in the neighboring ARG residues. (a) Distributions between CHB O11 - ARG H22 and CHB O15 - ARG HE in PDB: 3ZO8, with a representative molecular visualization in (b). (c) Distributions between CHB O1 - ARG H21 and CHB O2 - ARG H12 in 2CHT, with a representative molecular visualization in (d).}
    \label{fig:cm_hb}
\end{figure}

\begin{figure}[ht!]
    \centering

    \begin{minipage}[t]{0.48\linewidth}
        \centering
        \includegraphics[
            width=\linewidth,
            height=0.48\textheight,
            keepaspectratio
        ]{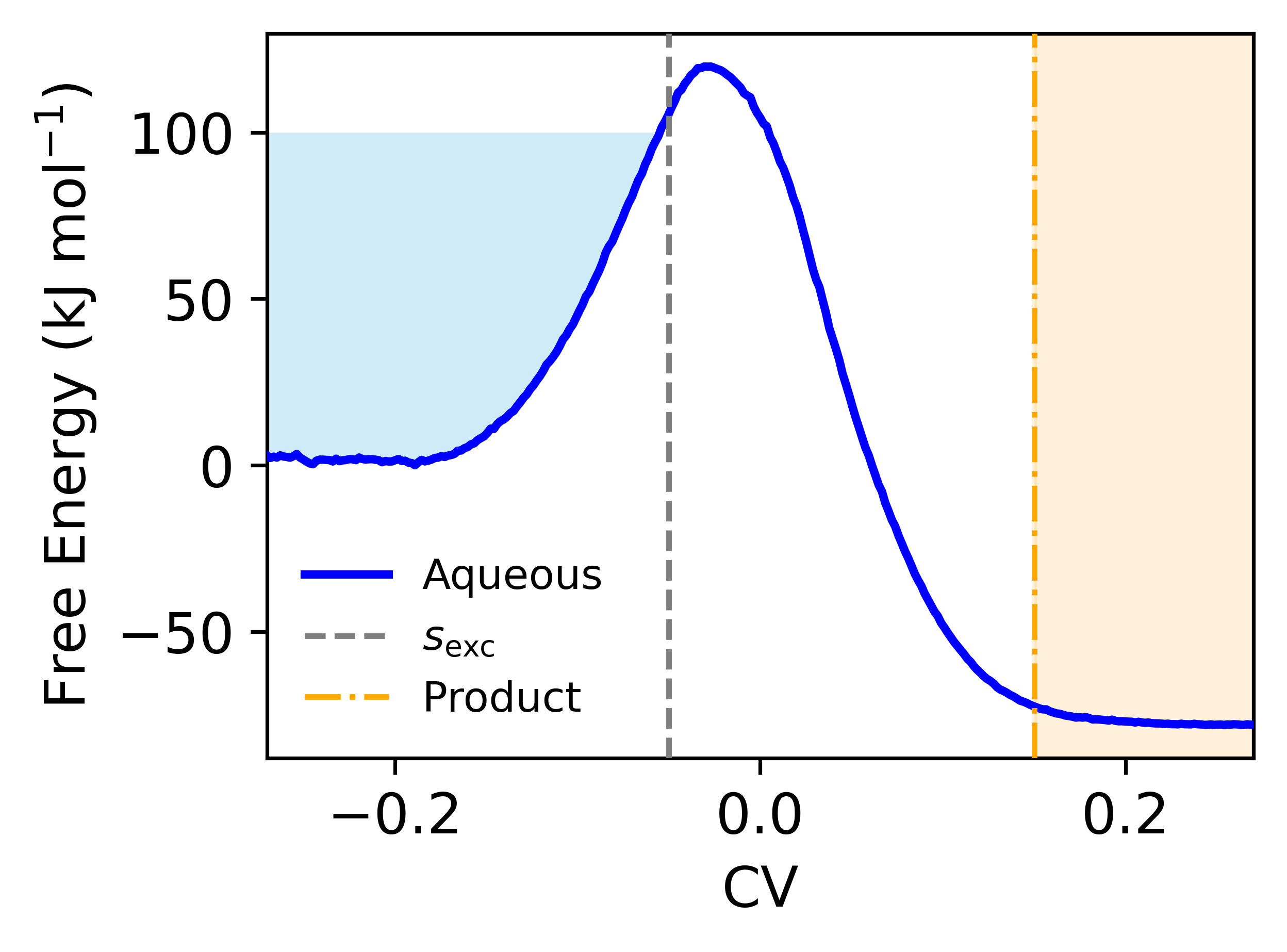}
    \end{minipage}
    \hfill
    \begin{minipage}[t]{0.48\linewidth}
        \centering
        \includegraphics[
            width=\linewidth,
            height=0.48\textheight,
            keepaspectratio
        ]{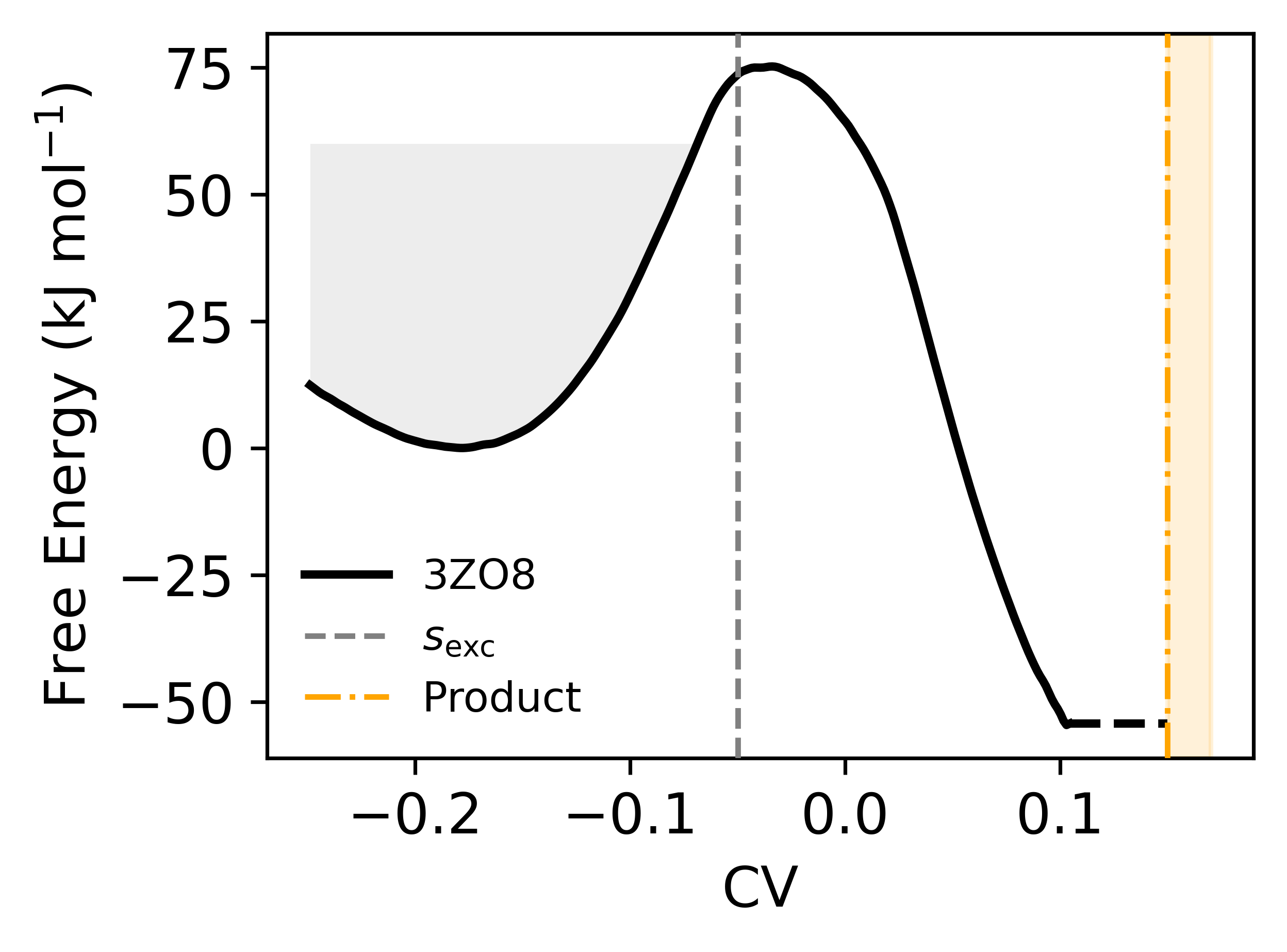}
    \end{minipage}

    \vspace{0.5cm}

    \begin{minipage}[t]{0.48\linewidth}
        \centering
        \includegraphics[
            width=\linewidth,
            height=0.28\textheight,
            keepaspectratio
        ]{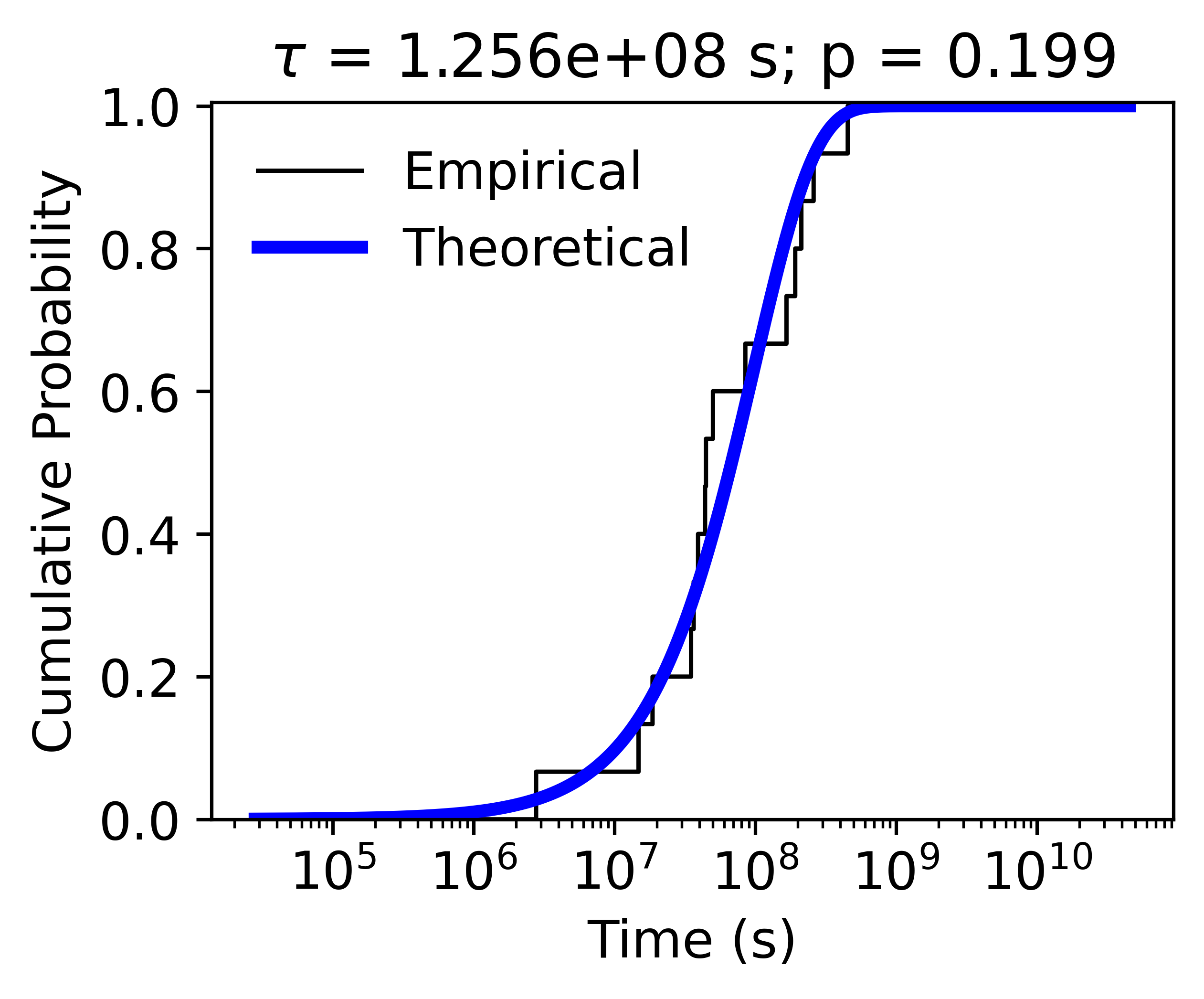}
    \end{minipage}
    \hfill
    \begin{minipage}[t]{0.48\linewidth}
        \centering
        \includegraphics[
            width=\linewidth,
            height=0.28\textheight,
            keepaspectratio
        ]{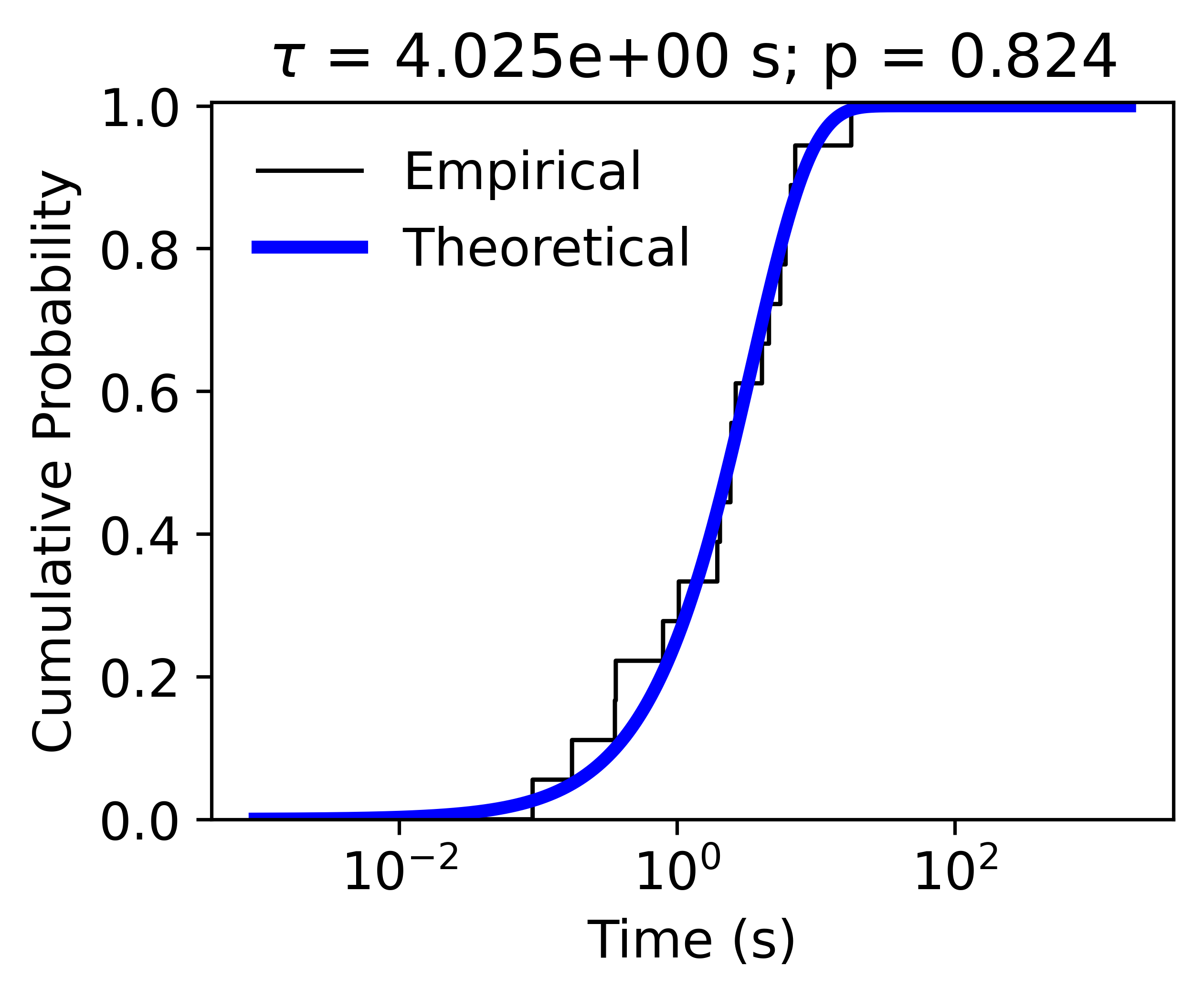}
    \end{minipage}

    \caption{(Top) \ourmodel-OPESf setup for computing the kinetics of Claisen rearrangement in an aqueous environment (left) and chorismate mutase 3ZO8 (right). (Bottom) Cumulative distribution of reweighted OPESf escape times fitted to a Poisson distribution obtained in the aqueous (left) and enzyme environments (right), respectively. The Poisson fits cannot be rejected under a two-sample KS test at the $\alpha$ = 0.05 significance level.}
    \label{fig:cm_opesf}
\end{figure}

\begin{figure}[ht!]
\centering
    \includegraphics[width=0.6\textwidth]{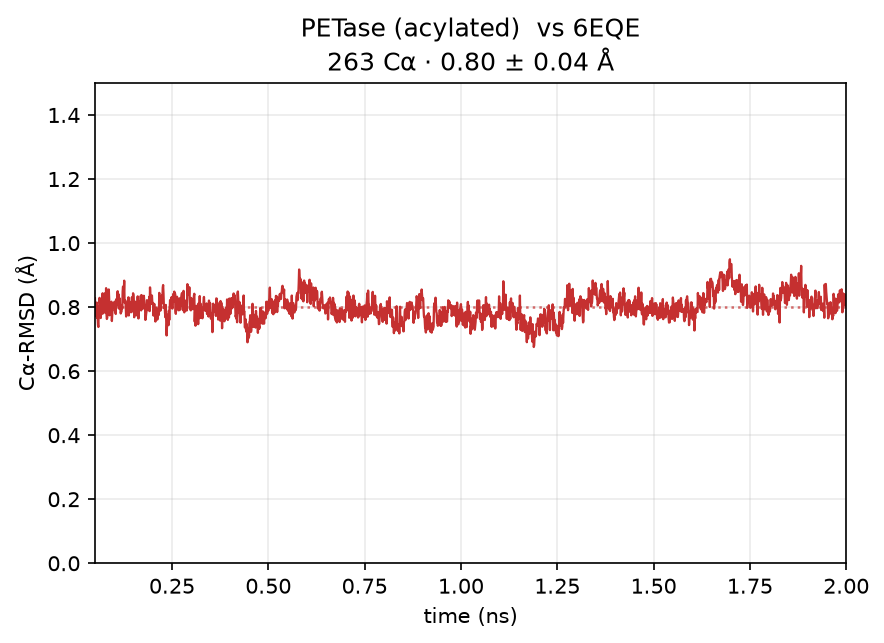}
    \caption{Trajectories of the PETase protein observed during an unbiased 2 ns NVT simulation. We observe that \ourmodel maintains a stable protein structure, with a $C\alpha$ RMSD of ($0.80 \pm 0.04$) \AA\ calculated relative to PDB: 6EQE \cite{pdb6eqe}.}
    \label{fig:petase_rmsd}
\end{figure}

We have quantified the error in barrier height predicted by \ourmodel, relative to its own reference level of theory ($\omega$B97M-V/def2-TZVPD).  The results are shown in Fig.~\ref{fig:cm_dft} and display excellent agreement.  To quantify the error in the barrier height relative to reference quantum chemistry (gold-standard) estimates, we have also recomputed the barrier with DLPNO-CCSD(T)/def2-QZVPPD using the geometry from $\omega$B97M-V/def2-TZVPD; the results are shown in Table~\ref{tab:chorismate_barriers}.  The \emph{ab initio} calculations were performed with ORCA version 6.0 and used $\omega$B97M-V/def2-TZVPD settings identical to the OMol25 dataset~\cite{levine_open_2025}.  Concretely, these calculations use the $\omega$B97M-V functional in conjunction with the def2-TZVPD basis set, the resolution of identity (RI) approximation and tight SCF convergence settings.  In addition to the SCF settings employed in the $\omega$B97M-V/def2-TZVPD calculations we use the TightPNO setting for DLPNO. The reactant and product structures were optimized at the $\omega$B97M-V/def2-TZVPD level; subsequently the TS geometry was obtained using the growing string method and optimized with ORCA's TS Opt. The transition state structure was verified at $\omega$B97M-V/def2-TZVPD level to contain one imaginary frequency.

\begin{table*}[t]
\centering
\caption{Gas phase barrier heights and reaction energies for the reaction chorismate-to-prephenate.  Geometries were optimized with $\omega$B97M-V/def2-TZVPD and single points evaluated with DLPNO-CCSD(T)/def2-QZVPPD.}
\label{tab:chorismate_barriers}
\begin{tabular}{lcc}
\toprule
Level of theory & Barrier height& Reaction energy \\
 & (kj/mol) & (kj/mol) \\
\midrule
$\omega$B97M-V/def2-TZVPD & 169.8 & -90.5\\
\ourmodel single-points & 170.5 & -89.0\\
\ourmodel relaxations & 169.0 & -89.0\\
DLPNO-CCSD(T)/def2-TZVPPD & 156.2 & -91.0 \\
DLPNO-CCSD(T)/def2-QZVPPD & 159.4 & -89.8\\
\bottomrule
\end{tabular}%
\end{table*}

\clearpage
\newpage
\section{PETase}\label{sec:petase_appendix}

\subsection{Simulation Setup and Details}\label{methods:petase}
Initial structures for both the acylation and deacylation simulations were obtained from \cite{burgin2024}. The deacylation simulations were initiated from the same acyl-enzyme intermediate (AEI) structure used in \cite{burgin2024}. In all the PETase simulations, we employed ASE \cite{ase-paper} together with \ourmodel \cite{wood_uma_2025}. Prior to the WTMetaD simulations, each system was first energy minimized and equilibrated. For energy minimization, we used a force tolerance of 0.05 eV/\AA. We performed a two-stage equilibration to gradually bring the system to the target temperature and density. In the first stage, the system was equilibrated in the NVT ensemble at $T=310$ K. Subsequently, the system was equilibrated in the NPT ensemble at $T=310$ K and $P=1$ bar. We employed the Langevin thermostat and IsotropicMTKNPT barostat during the equilibration and final scale simulations. 

\subsection{Reaction Coordinates}\label{petase_rc}
For all simulations in the PETase section, we used identical Reaction Coordinates (RCs) reported by \cite{burgin2024} for both the acylation and deacylation reactions. The acylation reaction coordinate (RC$_{acylation}$) is a linear combination of three collective variables (CV1a, CV2a, CV3a). CV1a and CV2a correspond to the distance differences $d_{SER160:H-PET:O}-d_{SER160:O-PET:C}$ and $d_{SER160:O-PET:C}-d_{PET:C-PET:O}$, respectively. CV3a corresponds to the angle formed by SER160 H--PET ether O--BHET C. Similarly, the deacylation reaction coordinate (RC$_{deacylation}$) is a linear combination of three collective variables (CV1d, CV2d, CV3d). CV1d, CV2d, and CV3d correspond to the distance differences $d_{SER160:O-Wat:H}-d_{Wat:H-Wat:O}$, $d_{SER160:O-MHET:C}-d_{MHET:C-Wat:O}$, and $d_{MHET:C-Wat:O}-d_{Wat:O-Wat:H}$, respectively. These reaction coordinates collectively describe the proton-transfer, nucleophilic attack, and bond-breaking/bond-forming events during the acylation and deacylation reactions. The RCs are:

\begin{align*}
    RC_{\text{acylation}} = 4.18 - 2.32 * \text{CV1a} - 2.59 * \text{CV2a} - 0.0119 * \text{CV3a}\\
    RC_{\text{deacylation}} = 1.55 - 1.42 * \text{CV1d} + 1.39 * \text{CV2d} - 0.590 * \text{CV3d} \\
\end{align*}

\subsection{Replica Exchange Umbrella Sampling}\label{sec:petase_reus}

For mapping the free energy surface of the full reaction during both acylation and deacylation, we used replica exchange umbrella sampling (REUS) \cite{sugita2000multidimensional}. In this variant of umbrella sampling (US) \cite{torrie1977nonphysical}, exchange of configurations between neighboring windows is allowed through a Metropolis criterion to improve convergence and sampling of orthogonal degrees of freedom. In particular, for a given window $i$, assuming a 1D CV $q(x)$, the total energy of the system in REUS is described with $H_{i}(x,p)=U(x)+K(p)+w_{i}(q(x))$ where $w_i(q)=\frac12 k_{i}(q-c_{i})^2$ represents a harmonic bias potential added to restrain and sample the system near the window center $c_{i}$ with a force constant of $k_{i}$. $H_{i}(x, p)$, $U(x)$, and $K(p)$ represent the Hamiltonian, potential energy, and kinetic energy of the system, respectively. $q$ and $p$ refer to the configuration and momentum of the system, respectively. The configurations and momenta from neighboring windows $i$ and $j$ in REUS are exchanged every 1000 simulation steps by using the Metropolis criterion with acceptance probability $P_{\rm acc}=\min[1,\exp(\beta\delta)]$ where $\delta = w_i(q_i)+w_j(q_j) -w_i(q_j)-w_j(q_i)$ and $\beta = \frac{1}{k_B T}$. When $\delta >= 0$, the swap lowers the overall energy and is always accepted. 

\begin{figure*}[ht!]
\centering
\includegraphics[width=\textwidth]{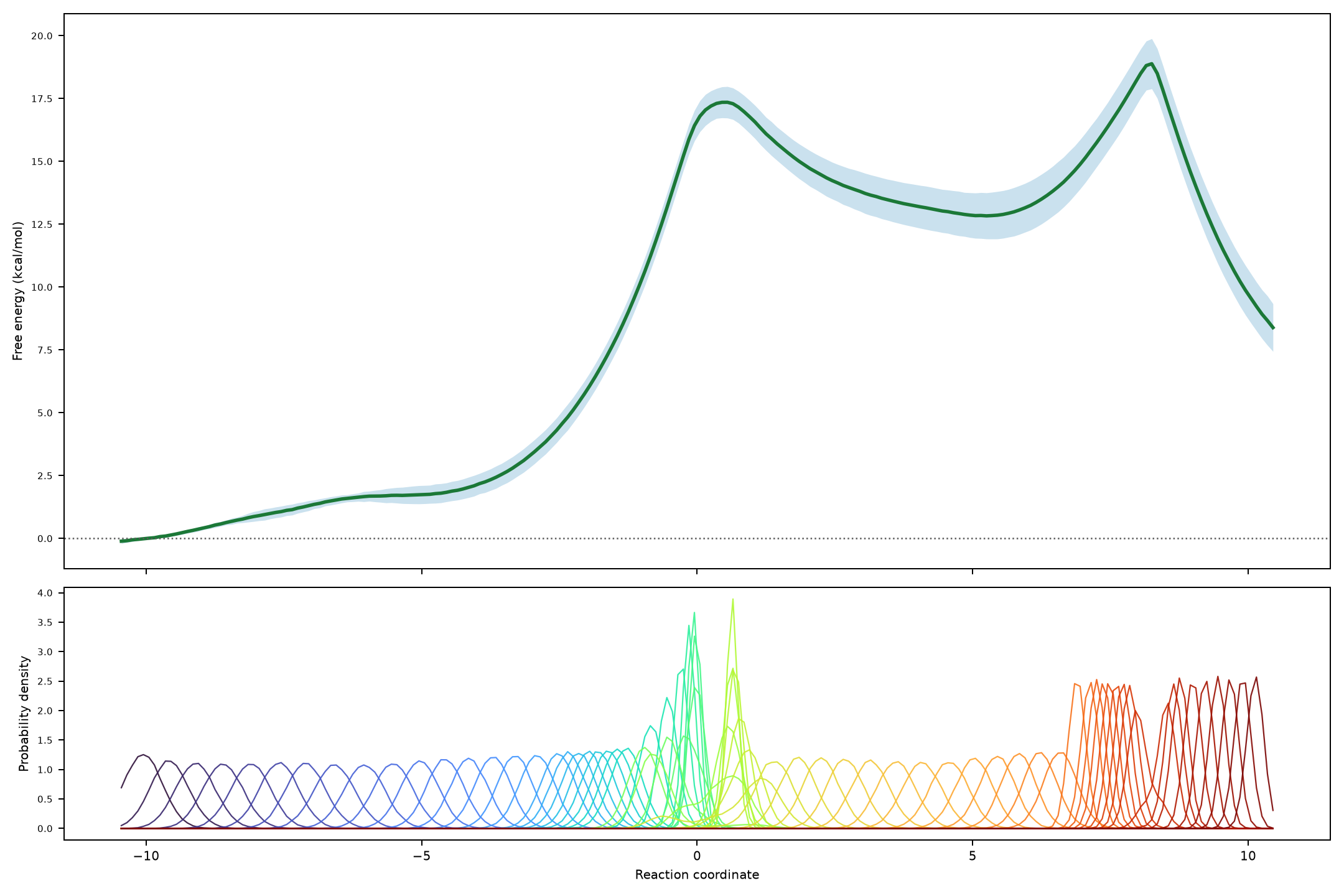}
\caption{BayesWHAM free-energy profile for PETase acylation using the curated 52-window connected umbrella set. The shaded band is the pointwise 95\% posterior interval.}
\label{fig:petase-acylation-overlap}
\end{figure*}

\begin{figure*}[ht!]
\centering
\includegraphics[width=\textwidth]{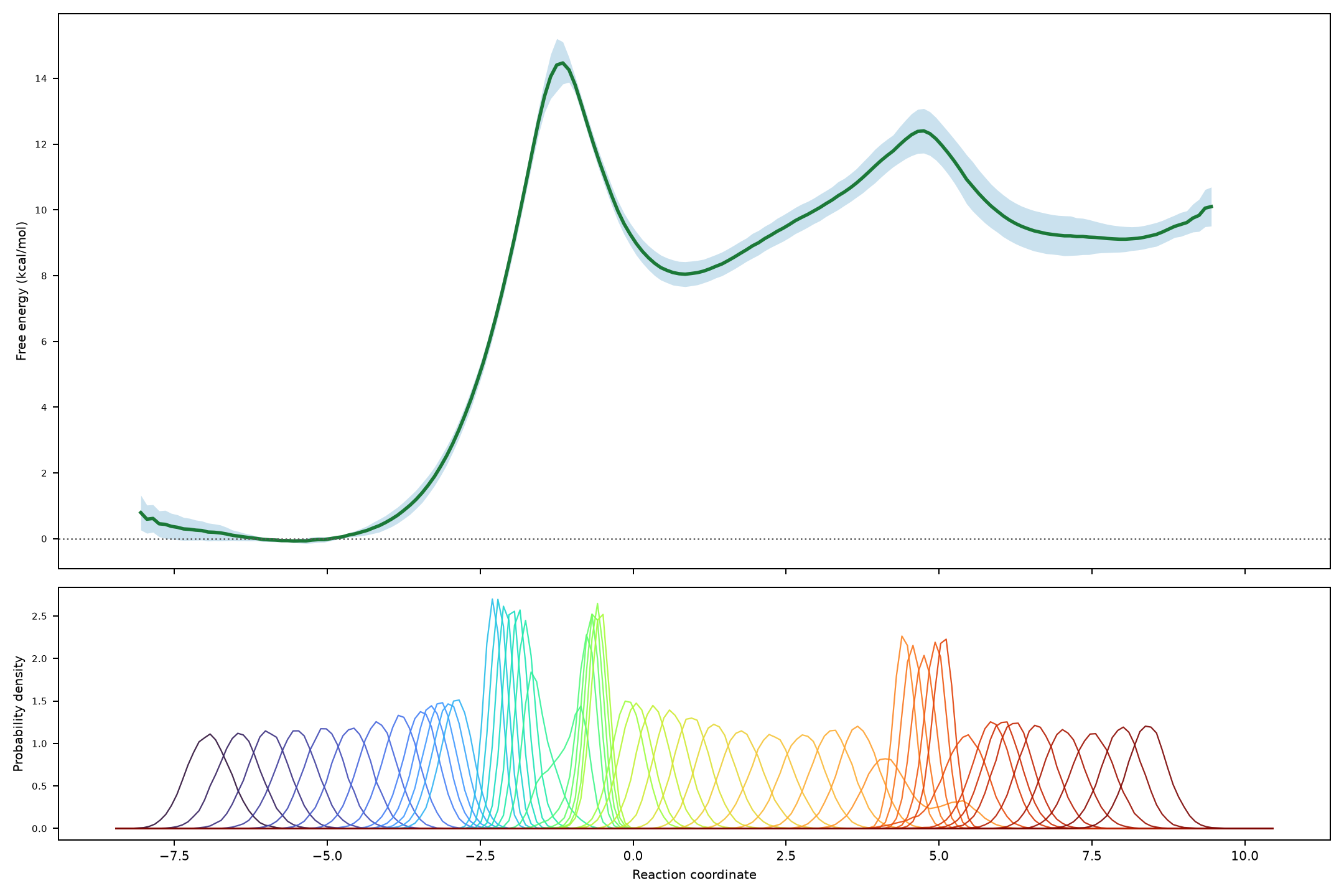}
\caption{BayesWHAM free-energy profile for PETase deacylation using the curated 50-window connected umbrella set. The shaded band is the pointwise 95\% posterior interval.}
\label{fig:petase-deacylation-overlap}
\end{figure*}

We obtain the starting configurations for each window by computing an approximate profile of the full free energy surface using OPES MetaD \cite{invernizzi_rethinking_2020} simulation by patching ASE with PLUMED \cite{plumed}. In Figures \ref{fig:petase-acylation-overlap} and \ref{fig:petase-deacylation-overlap}, we show the resulting histograms of $RC_{acylation}$ and $RC_{deacylation}$ in each window during the acylation and deacylation REUS runs, respectively. We adjusted window centers ($c_{i}$) and the value of the spring force constant ($k_{i}$) as needed to restrain the system in a given window and improve the overlap of the biased histograms of the neighboring windows in both $RC_{acylation}$ and $RC_{deacylation}$ as shown in 
Tables \ref{tab:petase-acylation-windows}
and \ref{tab:petase-deacylation-windows}, respectively. These tables also show the length of the simulation performed in each window. We compute the free energy profiles from the resulting biased distributions of $RC_{acylation}$ and $RC_{deacylation}$ in each window using BayesWHAM \cite{kumar1992weighted, bayeswham}. We detail the key features on the free energy surfaces in Table \ref{tab:petase-reus-features}. The total stable sampling time was $\sim$400 ps per window, totaling $\sim50$ ns of simulation time. The samples were divided into 6 even blocks to obtain measurement errors. 

\clearpage
\newpage

\begin{longtable}{rrr@{\qquad}rrr@{\qquad}rrr}
\caption{Umbrella centers, force constants, for the curated 74-window WT acylation analysis. Total 5.5M samples in 30 ns of sampling, $\sim$400 ps per window.}\label{tab:petase-acylation-windows} \\
\toprule
ID & Center & $k$ & ID & Center & $k$ & ID & Center & $k$ \\
\midrule
\endfirsthead
\toprule
ID & Center & $k$ & ID & Center & $k$ & ID & Center & $k$ \\
\midrule
\endhead
0 & -10 & 20 & 71 & -0.15 & 75 & 35 & 4.5 & 20 \\
1 & -9.5 & 20 & 72 & -0.1 & 150 & 36 & 5 & 20 \\
2 & -9 & 20 & 73 & -0.05 & 250 & 37 & 5.5 & 20 \\
3 & -8.5 & 20 & 69 & 0 & 300 & 38 & 6 & 20 \\
4 & -8 & 20 & 74 & 0.05 & 250 & 39 & 6.5 & 20 \\
5 & -7.5 & 20 & 75 & 0.1 & 150 & 40 & 7 & 20 \\
6 & -7 & 20 & 76 & 0.15 & 75 & 51 & 7 & 100 \\
7 & -6.5 & 20 & 77 & 0.2 & 40 & 52 & 7.25 & 100 \\
8 & -6 & 20 & 25 & 0.25 & 20 & 53 & 7.375 & 100 \\
9 & -5.5 & 20 & 26 & 0.5 & 20 & 54 & 7.5 & 100 \\
10 & -5 & 20 & 79 & 0.525 & 40 & 55 & 7.625 & 100 \\
11 & -4.5 & 20 & 80 & 0.55 & 75 & 56 & 7.75 & 100 \\
12 & -4 & 20 & 81 & 0.6 & 150 & 57 & 7.875 & 100 \\
13 & -3.5 & 20 & 78 & 0.625 & 300 & 58 & 8 & 100 \\
14 & -3 & 20 & 82 & 0.65 & 150 & 59 & 8.125 & 100 \\
15 & -2.5 & 20 & 83 & 0.7 & 75 & 60 & 8.25 & 100 \\
16 & -2 & 20 & 84 & 0.725 & 40 & 68 & 8.375 & 100 \\
17 & -1.75 & 20 & 27 & 0.75 & 20 & 61 & 8.5 & 100 \\
18 & -1.5 & 20 & 28 & 1 & 20 & 62 & 8.75 & 100 \\
19 & -1.25 & 20 & 29 & 1.5 & 20 & 63 & 9 & 100 \\
20 & -1 & 20 & 30 & 2 & 20 & 64 & 9.25 & 100 \\
21 & -0.75 & 20 & 31 & 2.5 & 20 & 65 & 9.5 & 100 \\
22 & -0.5 & 20 & 32 & 3 & 20 & 66 & 9.75 & 100 \\
23 & -0.25 & 20 & 33 & 3.5 & 20 & 67 & 10 & 100 \\
70 & -0.2 & 40 & 34 & 4 & 20 &  &  &  \\
\bottomrule
\end{longtable}

\begin{longtable}{rrr@{\qquad}rrr@{\qquad}rrr}
\caption{Umbrella centers, force constants, for the curated 52-window WT deacylation analysis. Total 3.7M samples in 21 ns of sampling, $\sim$400 ps per window.}\label{tab:petase-deacylation-windows} \\
\toprule
ID & Center & $k$ & ID & Center & $k$ & ID & Center & $k$ \\
\midrule
\endfirsthead
\toprule
ID & Center & $k$ & ID & Center & $k$ & ID & Center & $k$ \\
\midrule
\endhead
2 & -7 & 20 & 59 & -1.375 & 100 & 30 & 4 & 20 \\
3 & -6.5 & 20 & 60 & -1.25 & 100 & 31 & 4.5 & 20 \\
4 & -6 & 20 & 61 & -1.125 & 100 & 49 & 4.5 & 100 \\
5 & -5.5 & 20 & 44 & -1 & 100 & 50 & 4.625 & 100 \\
6 & -5 & 20 & 45 & -0.9375 & 100 & 51 & 4.75 & 100 \\
7 & -4.5 & 20 & 46 & -0.875 & 100 & 52 & 4.875 & 100 \\
8 & -4 & 20 & 47 & -0.8125 & 100 & 32 & 5 & 20 \\
9 & -3.5 & 20 & 20 & -0.75 & 20 & 53 & 5 & 100 \\
10 & -3 & 20 & 48 & -0.75 & 100 & 33 & 5.5 & 20 \\
11 & -2.75 & 20 & 21 & -0.5 & 20 & 34 & 5.75 & 20 \\
12 & -2.5 & 20 & 22 & 0 & 20 & 35 & 6 & 20 \\
13 & -2.25 & 20 & 23 & 0.5 & 20 & 36 & 6.5 & 20 \\
14 & -2 & 20 & 24 & 1 & 20 & 37 & 7 & 20 \\
54 & -2 & 100 & 25 & 1.5 & 20 & 38 & 7.5 & 20 \\
55 & -1.875 & 100 & 26 & 2 & 20 & 39 & 8 & 20 \\
56 & -1.75 & 100 & 27 & 2.5 & 20 & 40 & 8.5 & 20 \\
57 & -1.625 & 100 & 28 & 3 & 20 &  &  &  \\
58 & -1.5 & 100 & 29 & 3.5 & 20 &  &  &  \\
\bottomrule
\end{longtable}

\clearpage
\newpage

\begin{table*}[t]
\centering
\caption{PETase FES landmarks.}
\label{tab:petase-reus-features}
\resizebox{\textwidth}{!}{%
\begin{tabular}{llrrrrr}
\toprule
Reaction & Feature & Mean RC & $\Delta G$ (kcal mol$^{-1}$) & 95\% CI (kcal mol$^{-1}$) & $\Delta G$ (kJ mol$^{-1}$) & 95\% CI (kJ mol$^{-1}$) \\
\midrule
Acylation & First transition (TS1) & 0.50 & 17.39 & [16.82, 17.97] & 72.78 & [70.37, 75.19] \\
Acylation & Intermediate basin 1 (I1) & 5.20 & 12.82 & [11.90, 13.74] & 53.65 & [49.80, 57.49] \\
Acylation & Second transition (TS2) & 8.22 & 18.89 & [17.90, 19.88] & 79.03 & [74.87, 83.19] \\
Deacylation & Third transition (TS3) & -1.17 & 14.51 & [13.79, 15.23] & 60.72 & [57.70, 63.74] \\
Deacylation & Intermediate basin 2 (I2) & 0.85 & 8.04 & [7.67, 8.42] & 33.66 & [32.07, 35.25] \\
Deacylation & Fourth transition (TS4) & 4.73 & 12.41 & [11.73, 13.08] & 51.91 & [49.07, 54.74] \\
\bottomrule
\end{tabular}%
}
\end{table*}

\subsection{Well Tempered Metadynamics}

\begin{figure}[ht!]
    \centering
    \includegraphics[width=0.5\linewidth]{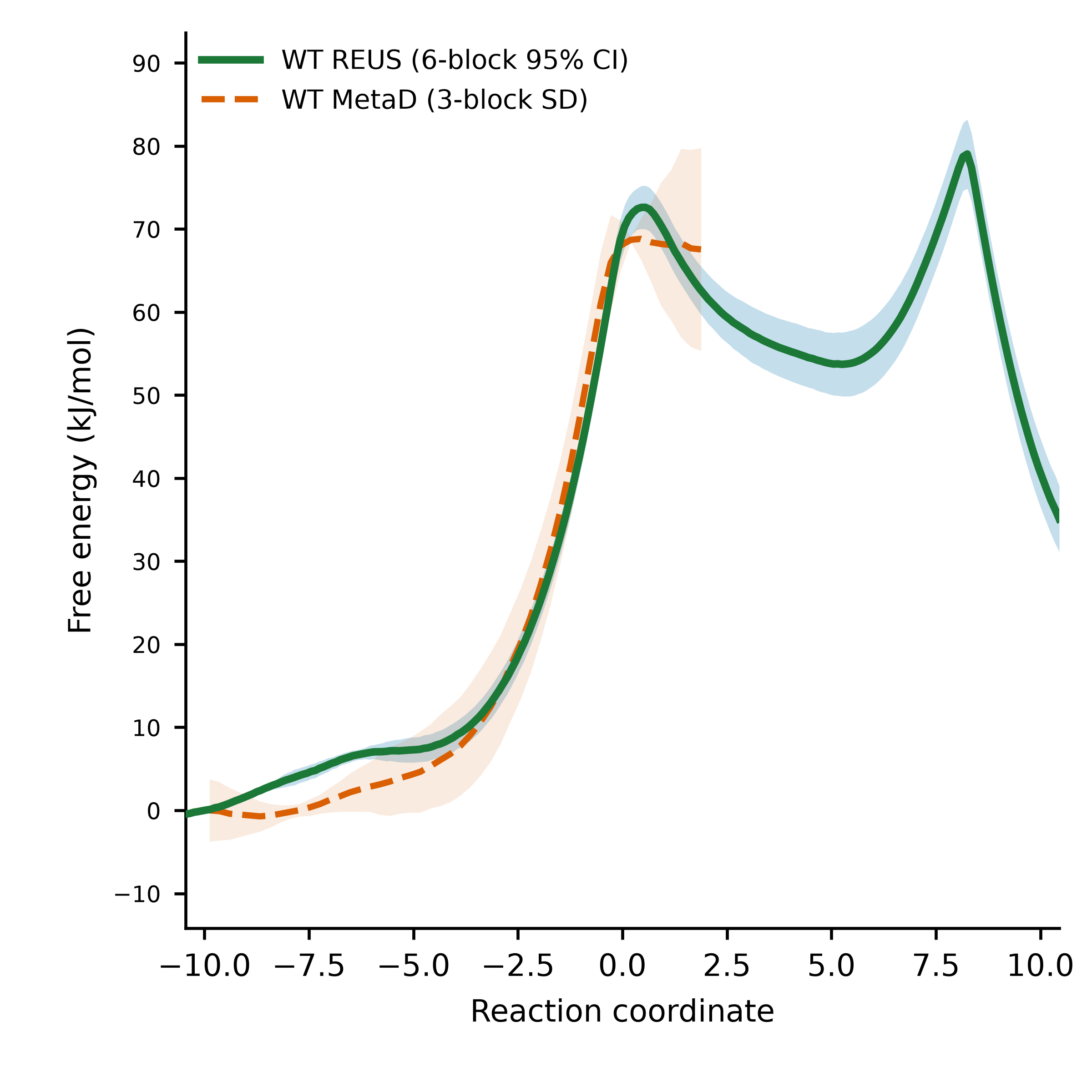}
    \caption{For acylation, we show the well-tempered metadynamics yields the first barrier height within error of REUS runs.}
    \label{fig:SI_pet_wt_acylation_reus_metad_overlay}
\end{figure}

\begin{figure}[ht!]
    \centering
    \includegraphics[width=0.5\linewidth]{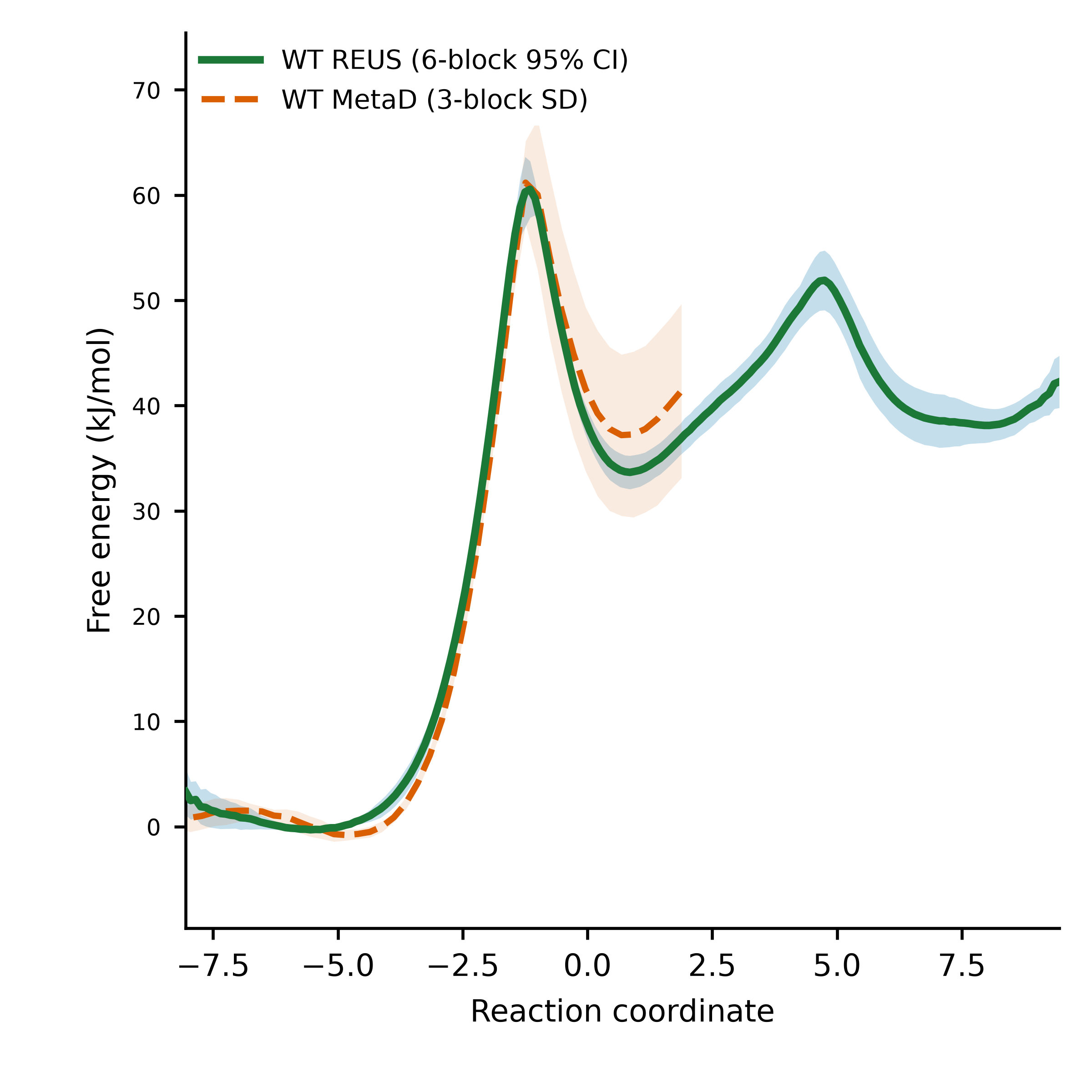}
    \caption{For deacylation, we show that the well-tempered metadynamics runs result in the first barrier height within error of REUS runs.}
    \label{fig:SI_pet_wt_deacylation_reus_metad_overlay}
\end{figure}

In addition to REUS, we performed experiments with well-tempered metadynamics to support our REUS results (Figures \ref{fig:SI_pet_wt_acylation_reus_metad_overlay} and \ref{fig:SI_pet_wt_deacylation_reus_metad_overlay}). We found barrier values within error of our REUS measurements. Because WT MetaD took much longer to converge for such a long RC trajectory, we added soft harmonic wall potentials along the reaction coordinate in both the acylation and deacylation simulations to limit the sampling of only the first transition state (TS1).
For the acylation simulations, lower and upper walls were placed at RC$_{acylation}=(-11.0)$ and $2.2$, respectively, with a force constant of 200 kJ mol$^{-1}$. Similarly, for the deacylation simulations, lower and upper walls were applied at RC$_{deacylation}=(-10.0)$ and $2.0$, respectively, using the same force constant. In addition, an upper harmonic wall was applied to the distance between the attacking water oxygen atom and the substrate carbonyl carbon ($d_{\mathrm{Wat:O-MHET:C}}$) at 0.35 nm with a force constant of 2000 kJ mol$^{-1}$ to prevent the catalytic water molecule from diffusing away from the active site during the deacylation simulations. 
We set the initial Gaussian hill height to 4 $\mathrm{kJ~mol^{-1}}$ and the bias factor to 20. The bias was deposited every 500 steps. WTMetaD simulations were performed until convergence, which happened within 5 ns for both the acylation and deacylation reactions.

\begin{figure}[ht!]
    \centering
    \includegraphics[width=\linewidth]{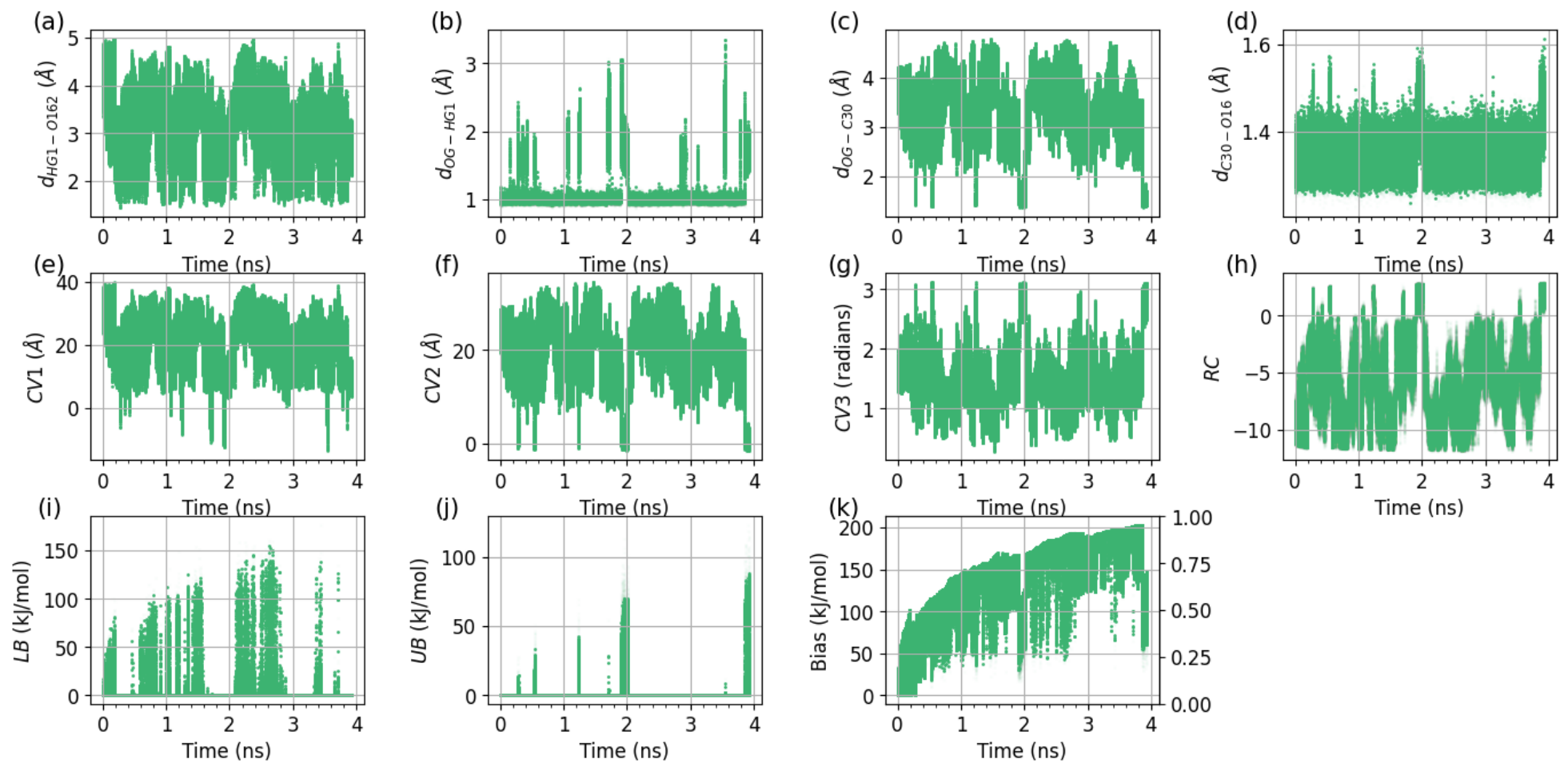}
    \caption{Convergence analysis of the WTMetaD simulation for the acylation reaction. Time evolution of (a) $d_{\mathrm{SER160:HG1-O16}}$, (b) $d_{\mathrm{SER160:OG-HG1}}$, (c) $d_{\mathrm{SER160:OG-C30}}$, (d) $d_{\mathrm{C30-O16}}$, (e) CV1a, (f) CV2a, (g) CV3, (h) RC$_{acylation}$, (i) the lower harmonic wall bias (LB), (j) the upper harmonic wall bias (UB), and (k) the accumulated WTMetaD bias.}
    \label{fig:SI_acylation_convergence}
\end{figure}

\begin{figure}[ht!]
    \centering
    \includegraphics{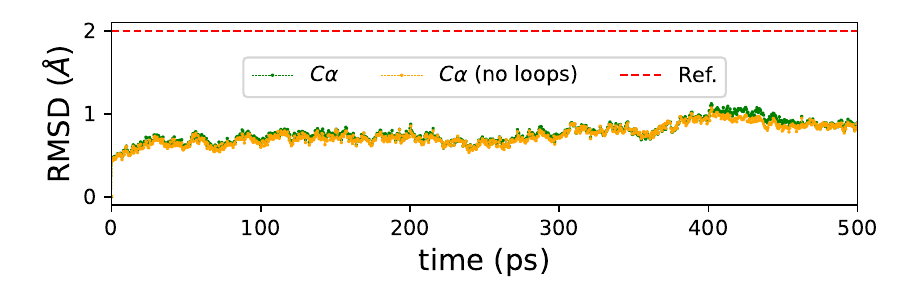}
    \caption{RMSD trajectory of 1KDN with bound ATP in water under NVT dynamics with \ourmodel. The simulation is stable and the protein remains folded with the C$_\alpha$ RMSD in sub-2 \AA\ agreement with the reference PDB: 1KDN throughout a 0.5 ns NVT simulation \cite{xu_alf3_1997}.
    }
    \label{fig:ndpk_rmsd}
\end{figure}

\Cref{fig:SI_acylation_convergence} shows the time evolution of the bias potential together with the sampling of all collective variables constituting RC$_{acylation}$. The accumulated bias plateaued within $\sim$5 ns, indicating convergence of the WTMetaD simulation within the sampled region. Furthermore, in both the acylation and deacylation simulations, the constituent CVs were well sampled throughout the region of interest.

\clearpage
\newpage

\clearpage
\newpage
\section{NDPK}\label{sec:NDPK_appendix}

\subsection{Simulation Background}
A conserved phosphohistidine intermediate forms within this active site, where a divalent cation (typically \ce{Mg^2+}) coordinates the nucleotide phosphates, balancing their negative charge and organizing the phosphoryl-transfer geometry. As such, accurate modeling of the dephosphorylation half-reaction is sensitive to the local electronic structure, cation coordination, and the local solvent context. The phosphoryl transfer step has proven difficult to model with conventional methods. Cluster models of the active site typically neglect the surrounding protein, ions, and solvent \cite{hutter2002mechanism}.

\subsection{Simulation Setup and Details}
Structurally, NDPK is a compact phosphoryl-transfer enzyme ($\sim$150 amino acids per subunit) that assembles into a homohexamer, with each subunit containing a conserved catalytic histidine. NDPK enables the conversion of ATP to ADP, preceding phosphoryl transfer to an incoming nucleoside diphosphate (NDP). High-resolution crystal structures of transition-state-like complexes such as PDB 1KDN have captured phosphoenzyme and transition-state analog geometries, providing unusually direct structural constraints on the catalytic mechanism. In PDB 1KDN, Dictyostelium NDPK is captured with Mg--ADP and aluminum fluoride, where AlF$_3$ mimics the transferring phosphoryl group \cite{xu_alf3_1997}. The AlF$_3$ moiety adopts a trigonal-planar arrangement that bridges the leaving-group oxygen of ADP (the $\beta$-phosphate O7) and the N$\delta$ of the catalytic histidine (HIS122), consistent with partial bonding to both the nucleophile and the leaving group in an in-line S$_\mathrm{N}$2-like transition state. 

The initial simulation box was prepared using the \texttt{CHARMM-GUI} input generator \cite{charmm-gui} by placing one NDPK hexamer with ATP docked inside the reactive pocket from the 1KDN \cite{1KDN} crystal structure within a cubic box with an edge length of 7 nm. We employed periodic boundary conditions in all dimensions.  The residues were protonated for a pH of 7 except for HIS122 where we enforced a doubly-protonated state as suggested by Hutter and Helms \cite{hutter2002mechanism}. The system was then solvated in water to a density of 1~g/cm$^3$. The total charge of the system was neutralized by adding a potassium ion. To generate the reactant state, consisting of the ATP substrate docked in the NDPK binding pocket, the AlF$_3$ residue in the 1KDN structure was replaced by a PO$_3$ residue and the system was energy minimized using the L-BFGS algorithm \cite{liu1989limited} until the maximum force was below 0.05~eV/\AA\ to eliminate high-energy overlaps. Equilibration was first performed for 100 ps in the NVT ensemble at 300 K using a Langevin thermostat with a friction constant of 1~ps$^{-1}$. Initial velocities were assigned from a Maxwell–Boltzmann distribution at 300~K. The classical equations of motion were integrated using a 0.5~fs time step. This was followed by 300~ps of equilibration in the NPT ensemble at 300~K and 1~bar using the isotropic Martyna-Tobias-Klein integrator \cite{martyna1994constant} with a temperature coupling time constant of 0.1~ps and a pressure coupling time constant of 1~ps. Final scale runs were conducted in the NVT ensemble at 300~K employing a Langevin integrator with a friction constant of 1~ps$^{-1}$. All MD simulations were conducted on 64 NVIDIA A-100 GPUs, yielding a performance of $\sim$1 ns/day.

In our simulations, the bound ATP adopts hydrogen-bonding contacts to TYR56, ARG92, THR98, and ASN119 in agreement with the 1KDN crystal structure\cite{1KDN} and prior semi-empirical results \cite{hutter2002mechanism}, and the Mg$^{2+}$ ion spontaneously adopts octahedral hexacoordination with the three phosphoryl groups of ATP, two water molecules, and GLU58 (\Cref{fig:ndpk}B). This geometry emerges without ion-specific empirical corrections. In contrast, MM approaches often fail to reproduce the Mg$^{2+}$ solvation free energy and first-shell geometry \cite{grotz2021optimized,panteva2015force} and QM/MM simulations require \textit{a priori} specification of the residues, solvent molecules, and ions participating in the QM region.

\subsection{Enhanced Sampling Calculations}
Enhanced sampling of the phosphoryl transfer reaction was conducted to study the reaction mechanism and estimate characteristic transition times. 
To understand the thermodynamics and the mechanism of the phosphotransfer reaction, we used two collective variables (CVs) which track nucleophilic attack of HIS and the proton transfer from the doubly-protonated HIS to the phosphoryl group. Both CVs were constructed using coordination numbers \cite{iannuzzi2003efficient}. We define the coordination number $S$ between the atom groups $A$ and $B$ as,
\begin{equation}
    S(n, m, r_0, d_0) = \sum_{i \in A}\sum_{j \in B} f(r_{ij};n, m, r_0, d_0),
\end{equation}
where,
\begin{equation}
    f(r_{ij}; n, m, r_0, d_0) = \frac{1- [(r_{ij}-d_0)/r_0]^n}{1- [(r_{ij}-d_0)/r_0]^m}.
\end{equation}
In the above definition, $d_0$ is the minimum inter-particle distance, $r_0$ is the midpoint of the switching function curve, and $n$ and $m$ are
non-negative numbers which control the steepness of the switching function curve.
We constructed the first collective variable, $\mathrm{CV}_1$, to track the phosphoryl transfer process as,
\begin{equation}
    \mathrm{CV}_1 = S_{\text{P-N}}(6, 12, 0.18, 0.0) - S_{\text{P-O}}(6, 12, 0.18, 0.0),
\end{equation}
where $S_{\text{P-N}}$ represents the coordination number of $\gamma$-P of the phosphoryl group with the acceptor N$\delta$ on HIS and $S_{\text{P-O}}$ refers to the coordination number of $\gamma$-P with all the oxygen atoms on the $\beta$-phosphate groups of ADP. Similarly, the second collective variable $\mathrm{CV}_2$ tracks the proton hopping process from HIS to the transferring $\gamma$-phosphoryl group,
\begin{equation}
    \mathrm{CV}_2 = S_{\mathrm{O\!-\!H}}(6, 12, 0.12, 0.0) - S_{\mathrm{N\!-\!H}}(6, 12, 0.12, 0.0),
\end{equation}
where $S_{\text{O-H}}$ corresponds to the coordination number of the proton on N$\delta$ of HIS with all oxygen atoms on the transferring $\gamma$-phosphoryl group, and $S_{\text{N-H}}$ represents the coordination number of the proton with the N$\delta$ atom on HIS. Note that the inclusion of all oxygen atoms in the $\beta$-phosphate groups of ADP in $S_{\text{P-O}}$ and all oxygen atoms in the transferring $\gamma$-phosphoryl group in $S_{\text{O-H}}$ automatically includes all degenerate states.

\begin{figure}[ht!]
    \centering
    \includegraphics[width=0.9\textwidth]{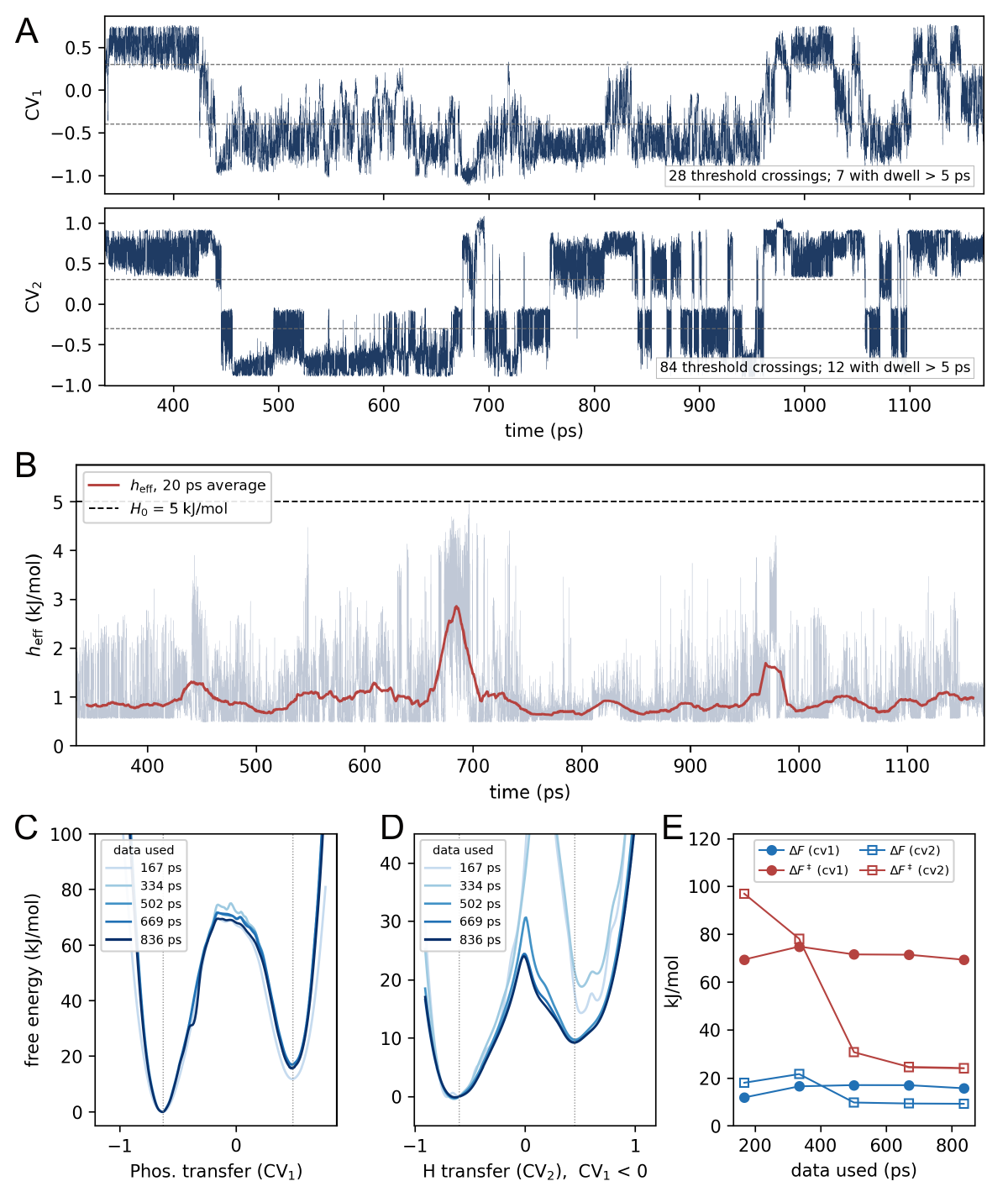}
    \caption{\textbf{(A)} Trajectories of the phosphoryl-transfer (CV$_1$) and proton-transfer (CV$_2$) coordinates during the WTMetaD simulation (first 20\% is not shown). Dashed lines mark the basin thresholds used to define transitions. The system diffuses between the reactant and product basins, giving 28 and 84 crossings in CV$_1$ and CV$_2$, respectively. Discarding rapid (< 5 ps) recrossings of the dividing surface leaves 7 and 12 independent basin exchanges. \textbf{(B)} Effective well-tempered deposition rate along the trajectory, $h_{\text{eff}} = H_0 \exp[-V(s)/k_B T(\gamma - 1)]$ with $k_B T(\gamma - 1)$ = 97.27 kJ/mol, shown raw (grey) and as a 20 ps running average (red); the dashed line is the initial hill height $H_0$ = 5 kJ/mol. Over the course of the simulation $h_{\text{eff}}$ is stationary at 0.8-1.1 kJ/mol, averaging 0.96 kJ/mol or 19\% of $H_0$. The bias had therefore essentially stopped growing, which justifies using the final, static bias as the reweighting potential. \textbf{(C-D)} Free energy along each coordinate computed from cumulative windows of increasing length, all beginning at 334 ps (\textbf{(C)}: CV$_1$, \textbf{(D)}: CV$_2$, conditioned on CV$_1$ < 0 as used for marginalization reported in Section \ref{app:ndpk_margin}, dotted lines mark the reference states). Both profiles are stationary after approximately 500 ps of data are included. \textbf{(E)} Over the three longest windows the reaction free energy varies by 0.8 kJ/mol (CV1) and 0.3 kJ/mol (CV2) and the barrier by 1.3 and 3.7 kJ/mol respectively. Cumulative rather than disjoint windows are used because individual basin residences last 28-285 ps, so a disjoint window can lie entirely within one basin and contain too few frames in the other to give a meaningful free-energy difference.
    }
    \label{fig:ndpk_metad_conv_app}
\end{figure}

The CVs defined above were used in a 2D WTMetaD simulation to study the free-energy landscape of the phosphotransfer reaction and extract mechanistic insight. During the WTMetaD simulations, a flat-bottom restraint was applied to the sum of CV$_1$ and CV$_2$ to prevent the proton from migrating away from the reaction site via water-mediated hops. A one-sided flat-bottom restraint was imposed between the center of mass of ATP and HIS122 to prevent the ATP substrate from leaving the active site. In practice, ATP remained docked throughout all calculations and the restraint never exerted a restoring force. A convergence assessment of the WTMetaD calculations is presented in Figure \ref{fig:ndpk_metad_conv_app}.

\begin{figure}[h!]
    \centering
    \includegraphics[width=0.9\textwidth]{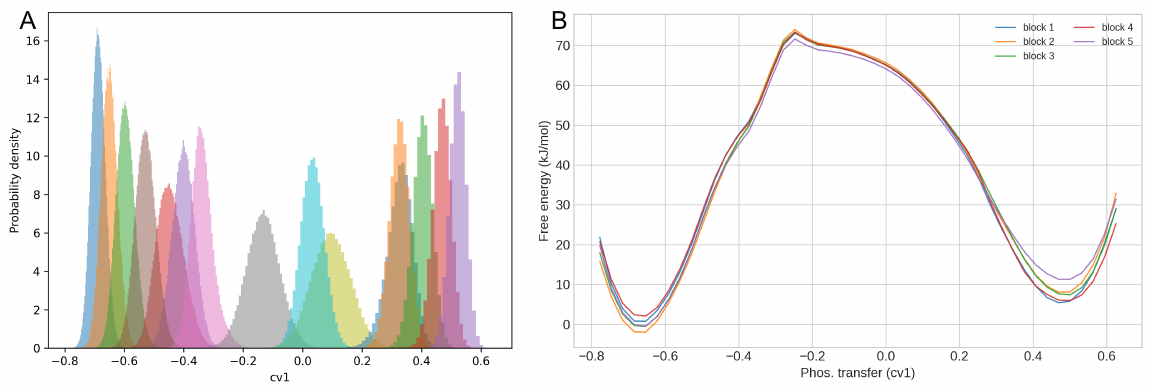}
    \caption{\textbf{(A)} The histogram of samples collected under each window for CV$_1$ shows sufficient overlap. \textbf{(B)} The PMFs calculated by splitting the US trajectories into five contiguous blocks are in good agreement.
    }
    \label{fig:ndpk_cv1_us_app}
\end{figure}

\begin{figure}[h!]
    \centering
    \includegraphics[width=0.9\textwidth]{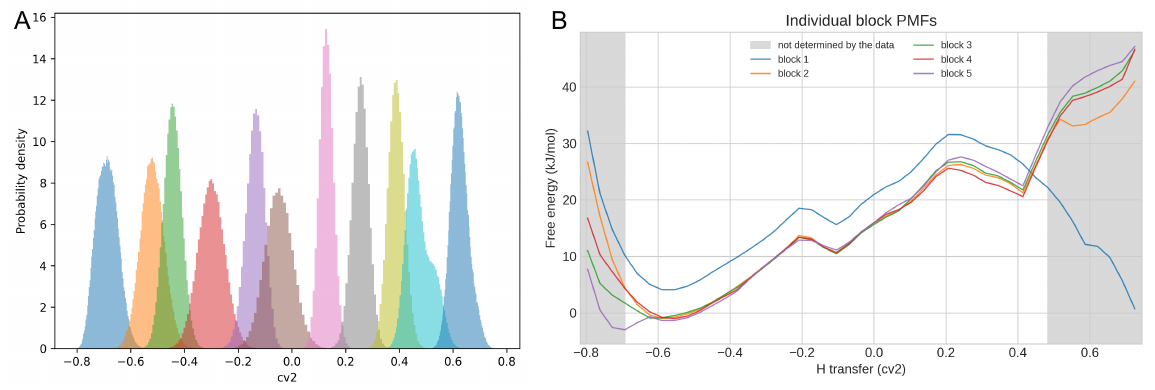}
    \caption{\textbf{(A)} The histogram of samples collected under each window for CV$_2$ shows sufficient overlap. \textbf{(B)} The PMFs calculated by splitting the US trajectories into five contiguous blocks are in good agreement.
    }
    \label{fig:ndpk_cv2_us_app}
\end{figure}

We further performed MD simulations with umbrella sampling along the phosphoryl transfer pathway using $\mathrm{CV}_1$ and along the proton transfer pathway using $\mathrm{CV}_2$ to improve the estimate of the location and height of the barrier along each path. The results of the MD simulations with umbrella sampling were reweighted using the BayesWHAM algorithm \cite{bayeswham} to calculate the potential of mean force (PMF) curve along $\mathrm{CV}_1$ and $\mathrm{CV}_2$ and their relative free energy. For the umbrella sampling runs on $\mathrm{CV}_1$, we use harmonic restraints with a force constant of 150.0 $\mathrm{kJ~mol^{-1}}$ to keep $\mathrm{CV}_2$ at 0.75. Umbrellas were placed along the phosphoryl-transfer coordinate and their force constants were tuned such that there was sufficient overlap between neighboring windows and the drift from each window was within acceptable range. \Cref{tab:umb1} shows the centers and force constants of the final umbrella setup. The samples collected from each window were then reweighted using the BayesWHAM algorithm \cite{bayeswham} to get the relative free energy of the windows and calculate the final PMF. A convergence assessment of these US calculations is presented in Figure \ref{fig:ndpk_cv1_us_app}.
We repeated this process with umbrellas along $\mathrm{CV}_2$, using harmonic restraints with a force constant of 150.0 $\mathrm{kJ~mol^{-1}}$ to keep $\mathrm{CV}_1$ at (-0.65), to get an estimate of the proton hopping barrier along the H transfer coordinate. \Cref{tab:umb2} shows the centers and force constants of the final umbrella setup. Similar to before, the samples collected from each window were then reweighted using the BayesWHAM algorithm \cite{bayeswham} to get the relative free energy of the windows and calculate the final PMF. A convergence assessment of these US calculations is presented in Figure \ref{fig:ndpk_cv2_us_app}.

\begin{table}[ht!]
\centering
\caption{Details of the umbrellas used for the phosphoryl transfer pathway.}\label{tab:umb1}
\begin{tabular}{|c|c|c|}
\hline
    Index & Center & Force const. ($\mathrm{kJ~mol^{-1}}$)\\
    \hline
  1 & -0.75 & 1000.00\\
  2 & -0.60 & 1000.00\\
  3 & -0.45 & 1000.00\\
  4 & -0.35 & 2000.00\\
  5 & -0.35 & 2750.00\\
  6 & -0.30 & 1000.00\\
  7 & -0.25 & 2500.00\\
  8 & -0.15 & 1000.00\\
  9 &  0.00 & 1000.00\\
 10 &  0.00 & 2000.00\\
 11 &  0.15 & 1000.00\\
 12 &  0.25 & 2000.00\\
 13 &  0.30 & 1000.00\\
 14 &  0.45 & 1000.00\\
 15 &  0.60 & 1000.00\\
 \hline
\end{tabular}
\end{table}

\begin{table}[ht!]
\centering
\caption{Summary of the Umbrellas used for the proton transfer pathway.}\label{tab:umb2}
\begin{tabular}{|c|c|c|}
\hline
    Index & Center & Force const. ($\mathrm{kJ~mol^{-1}}$)\\
    \hline
  1 & -0.750 & 1000.00\\
  2 & -0.500 & 1000.00\\
  3 & -0.425 & 2000.00\\
  4 & -0.250 & 1000.00\\
  5 & -0.150 & 2000.00\\
  6 &  0.000 & 1000.00\\
  7 &  0.150 & 3000.00\\
  8 &  0.250 & 3000.00\\
  9 &  0.375 & 3000.00\\
 10 &  0.500 & 3000.00\\
 \hline
\end{tabular}
\end{table}

\clearpage
\newpage

\clearpage
\newpage

\clearpage

\subsection{Marginalization of the Reweighted WTMetaD Samples}\label{app:ndpk_margin}

\begin{figure}[ht!]
    \centering
    \includegraphics{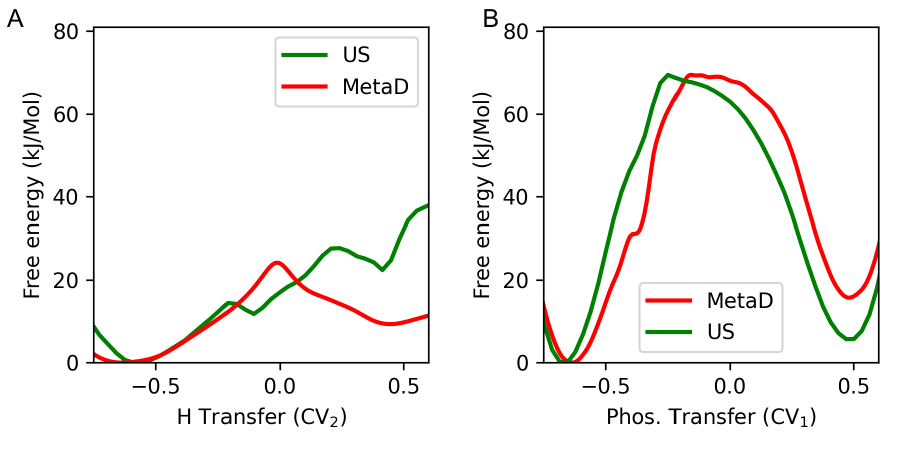}
    \caption{1D PMFs along CV$_2$ (H transfer) and CV$_1$ (phosphoryl transfer) from 1D umbrella sampling (US) superposed on those from marginalization of the results of the 2D WTMetaD. The PMFs for the two reactive processes computed under the two methods show good agreement in the location and height of the barrier, although the product along CV$_2$ is less stable in the US runs. This is likely a result of the marginalization process, where we only considered the samples with CV$_1$ values smaller than (-0.3) to isolate the destabilizing effect of the transition state along CV$_1$ which is not allowed to be visited in the umbrella sampling windows along CV$_2$.
    }
    \label{fig:ndpk_margin}
\end{figure}

To compare the FES obtained from the 2D WTMetaD run with the umbrella sampling results, we marginalized the 2D surface along CV$_1$ and CV$_2$. Figure \ref{fig:ndpk_margin} shows the marginalized PMFs superimposed on the umbrella sampling results. The location and height of the barriers along both CVs are in good agreement. We observe that the product of the H transfer reaction along CV$_2$ is at a higher free energy from the umbrella sampling results. We attribute this to the difference between how the product region is constrained between the two PMFs: in the umbrella sampling runs, a restraint on the P-O bond prevents the formation of the final product, whereas in the marginalized results the samples collected for CV$_1 > (-0.3)$ are excluded to isolate the destabilizing effect of the transition state as well as the final product along CV$_1$. Note that the PMF from US shown here is calculated without block-averaging.

\clearpage

\subsection{Cumulative Distribution of Reweighted OPESf Simulations for NDPK}

\begin{figure}[h!]
    \centering
    \includegraphics[width=0.5\linewidth]{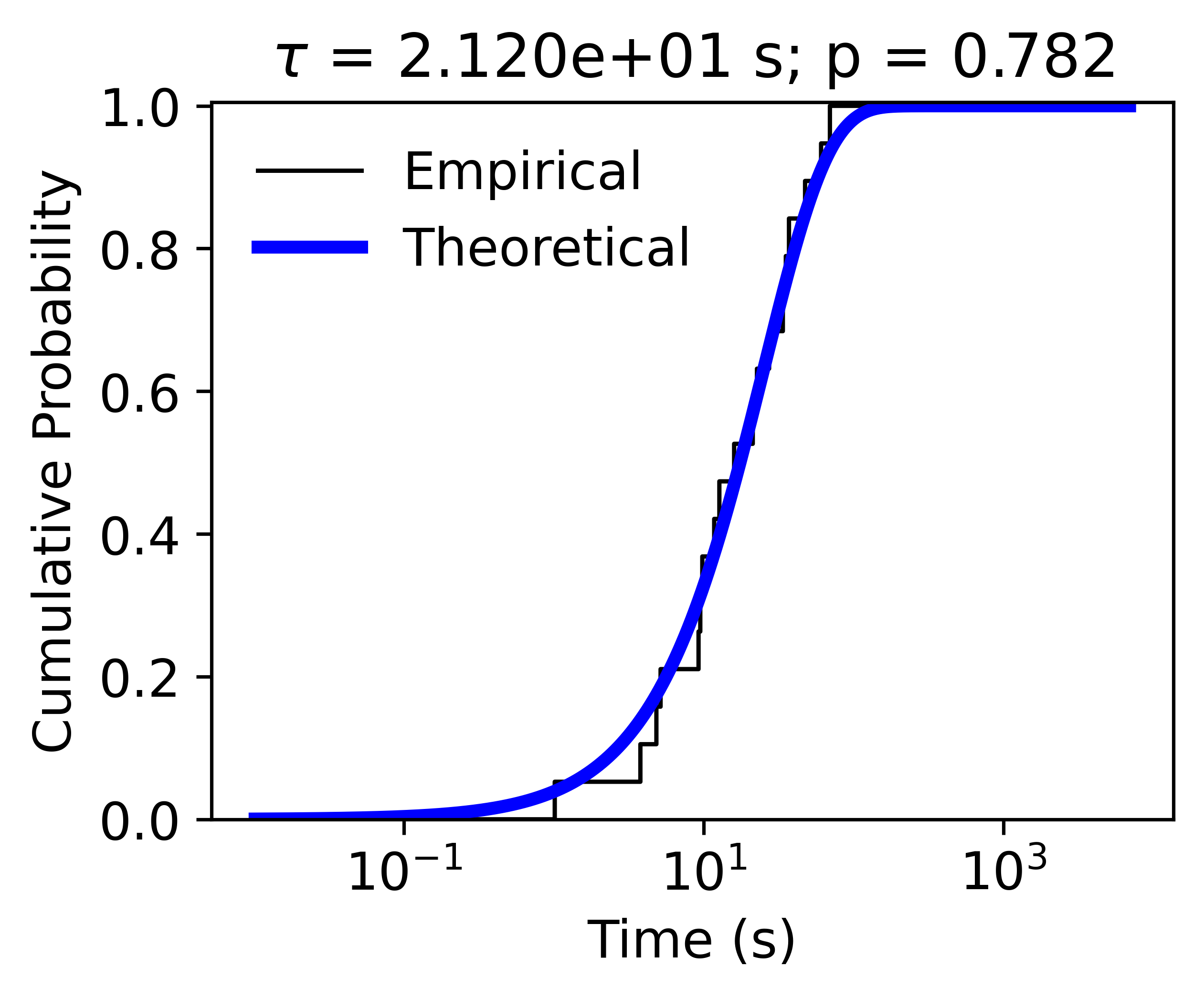}
    \caption{Cumulative distribution of reweighted OPESf phosphoryl transfer step escape times in NDPK fitted to a Poisson distribution. The Poisson fit cannot be rejected under a two-sample KS test at the $\alpha$ = 0.05 significance level.}
    \label{fig:ks_npdk}
\end{figure}

We applied OPESf to the rate limiting phosphoryl transfer step to calculate rate constants and effective activation energies using Eyring transition-state theory (\Cref{fig:ndpk}E, \Cref{sec:rate_eqn}) \cite{ray2022rare,seal_computing_2025}. Reweighted first-passage times from 19 independent OPESf trajectories follow Poisson statistics, yielding a predicted rate constant of $k_\text{OPESf}$ = (0.043 $\pm$ 0.019)~$\mathrm{s^{-1}}$ with a corresponding apparent effective activation free energy of $E_a^\text{OPESf}$ = (81.32 $\pm$ 1.13)~$\mathrm{kJ~mol^{-1}}$ (\Cref{fig:ks_npdk}).

\clearpage
\newpage
\section{Water Structure and Dynamics}\label{app:water}

We evaluated \ourmodel's capabilities in predicting the temperature dependence of the structural, thermodynamic, and kinetic properties of bulk water. An initial box of water containing 216 molecules was generated using GROMACS~\cite{lemkul2024introductory} and geometrically relaxed with the SPC/E water model~\cite{berendsen1987grigera}. Molecular dynamics simulations with \ourmodel were conducted using the integrators implemented in ASE~\cite{ase-paper}. All calculations were performed at neutral charge and in the singlet spin state.

\begin{figure}[ht!]
    \centering
    \includegraphics[width=0.75\textwidth]{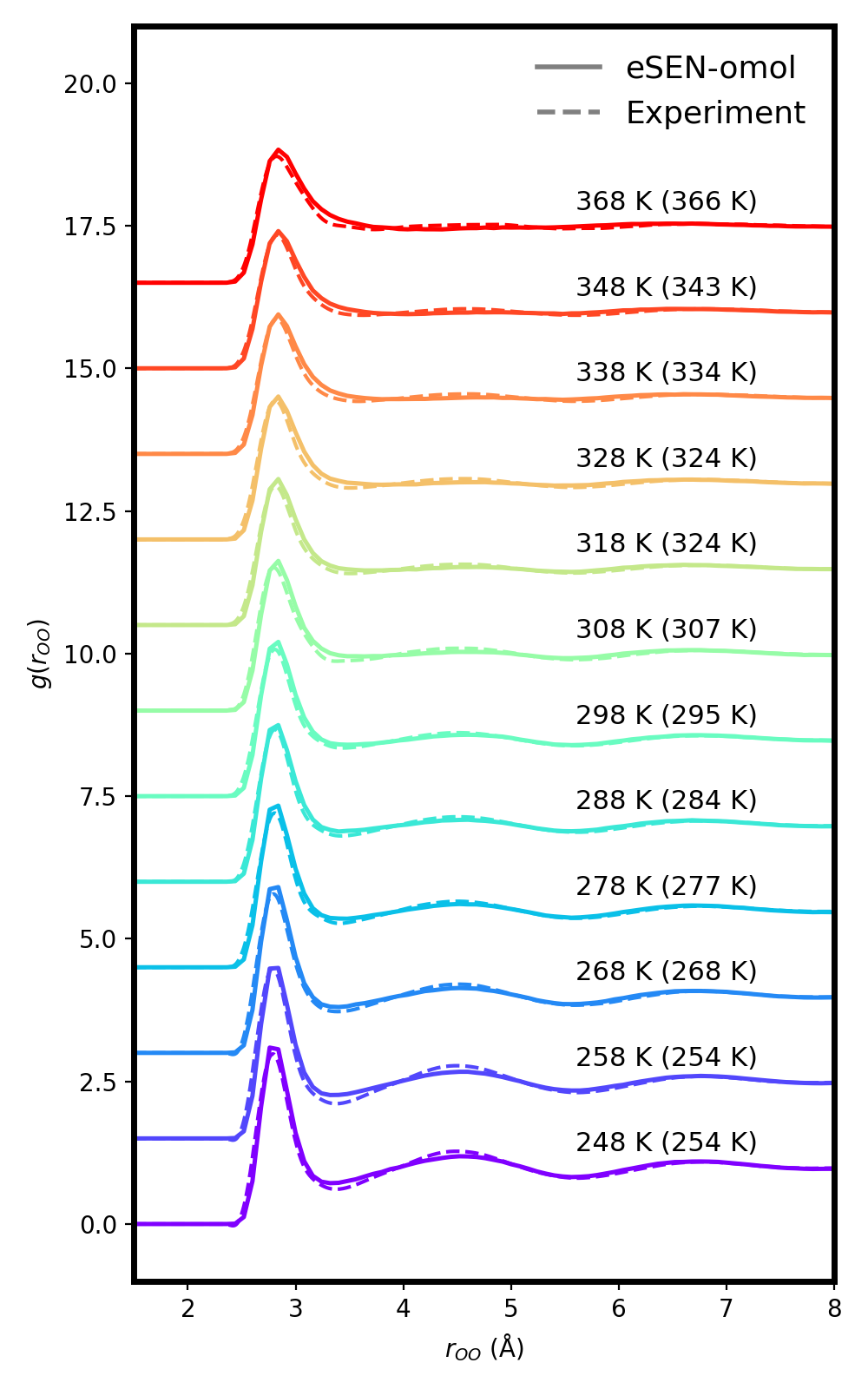}
    \caption{Temperature dependence of the oxygen-oxygen RDF of liquid water predicted by classical NVT simulations with \ourmodel compared with the corresponding results derived from X-ray diffraction measurements~\cite{skinner2013benchmark,skinner2014structure}. The
    temperatures at which the experimental measurements were performed are given in parentheses.}
    \label{water_rdf}
\end{figure}

We first computed radial distribution functions (RDFs) as a function of temperature to assess the structural behavior of \ourmodel for liquid water. These RDFs were calculated from equilibrated trajectories of 250~ps in the NVT ensemble using the Langevin integrator with a low friction coefficient of 0.001~fs$^{-1}$ and a time step of 0.5~fs, employing the \ourmodel model. Twelve independent simulations were performed at temperatures ranging from 248~K to 368~K. The resulting RDFs were compared against experimental X-ray diffraction measurements, for which we observe very good agreement~\cite{skinner2013benchmark,skinner2014structure} (\Cref{water_rdf}). 

\begin{figure}[ht!]
    \centering
    \includegraphics[width=0.75\textwidth]{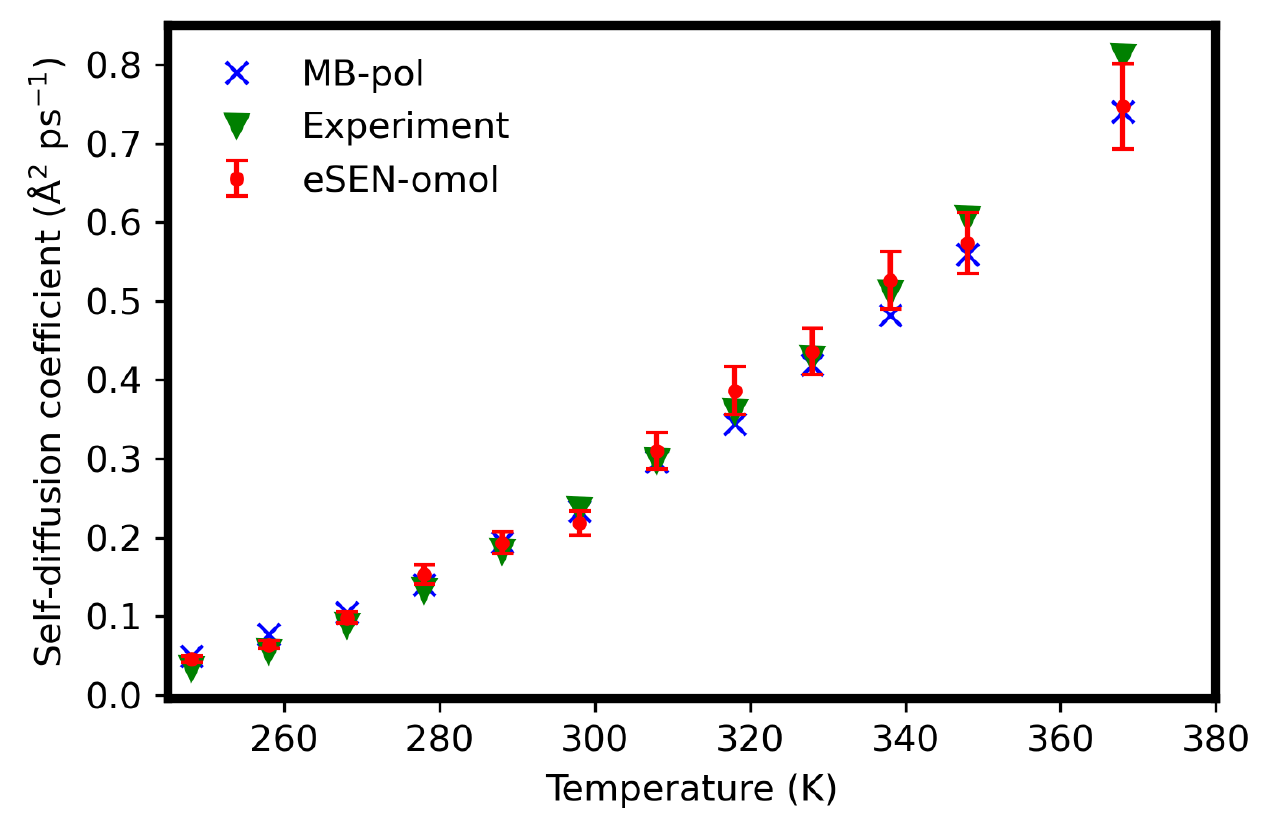}
    \caption{Self-diffusion coefficient of liquid water computed by \ourmodel, MB-pol \cite{reddy2016accuracy} and experimental measurements~\cite{gillen1972self,holz2000temperature,easteal1989diaphragm,mills1973self}.}
    \label{water_ds}
\end{figure}

We next used these simulation trajectories to evaluate the self-diffusion coefficient, $D_s$, from the mean-squared displacement (MSD) using the Einstein relation, $D_s = \lim_{t\to\infty}
          \frac{\langle |\mathbf{r}(t) - \mathbf{r}(0)|^2 \rangle}{6t},$

where the MSD was computed independently for each oxygen atom and averaged over all 216 atoms, with a linear fit applied in the diffusive regime (lag times 2--50~ps). It is well known that simulations employing periodic boundary conditions systematically underestimate the self-diffusion coefficient because a molecule couples hydrodynamically to its own periodic images through the long-ranged Oseen flow field~\cite{yeh2004system}. We applied the Yeh--Hummer finite-size correction, $D_\infty = D_\mathrm{PBC} + (k_B T\,\xi)/(6\pi\,\eta\,L),$

where $\xi = 2.837297$ is a dimensionless geometric constant for a cubic periodic box, $L$ is the box side length, and $\eta$ is the shear viscosity of water. Reference viscosity values were taken from the IAPWS international formulation for $T \geq 273$~K~\cite{huber2009viscosity} and from microfluidic Brownian-motion measurements in the supercooled regime for $T < 273$~K~\cite{dehaoui2015viscosity}. The corrected diffusivities predicted by \ourmodel are in excellent agreement with MB-pol \cite{reddy2016accuracy} and experimental measurements~\cite{gillen1972self,holz2000temperature,easteal1989diaphragm,mills1973self} (\Cref{water_ds}). 

\begin{figure}[ht!]
    \centering
    \includegraphics[width=0.75\textwidth]{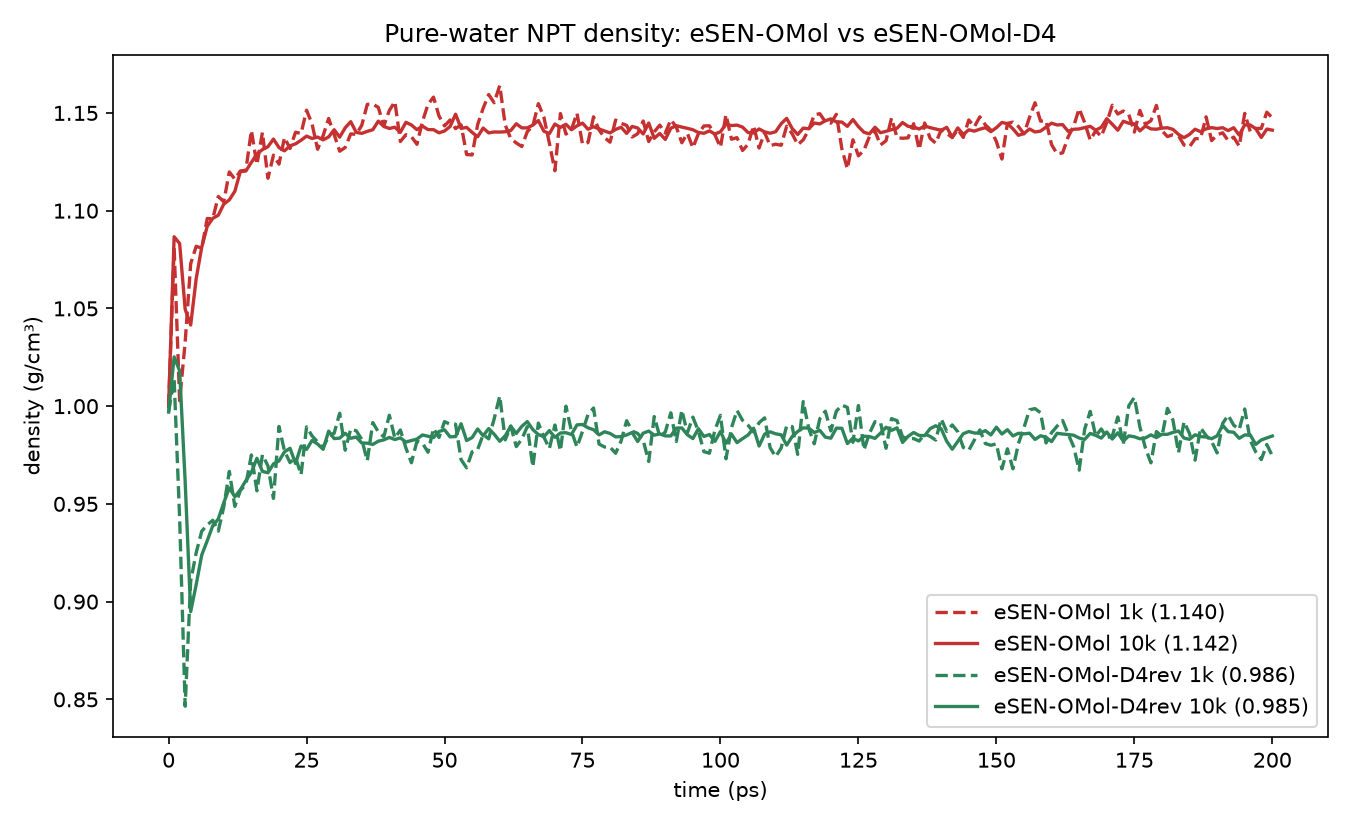}
    \caption{Water densities. \ourmodel produces higher than expected water density at 1~atm due to the neglect of 3-body dispersion in the VV10 functional used for training data \cite{levine_open_2025}. Finetuning \ourmodel on 4M DFT-D4Rev \cite{friede_optimally_2023} calculations recovers the correct density. Densities are invariant to the number of waters and box sizes (e.g.: 1k and 10k).}
    \label{water_density_d4}
\end{figure}

Finally, we evaluated the density of liquid water at 300K and 1 atm pressure. We generated boxes of water with TIP3P geometry using Open-MM (1k and 10k waters) and ran isotropic Martyna–Tobias–Klein~\cite{martyna1994constant} with a temperature damping of 100~fs and a pressure damping of 1000~fs for 200 ps each with 0.5 fs time steps. \ourmodel, like all other OMol-trained models, is known to have an elevated water density due to the neglect of many-body dispersion in the VV10 functional used as part of the $\omega$B97M-V functional in OMol25 \cite{levine_open_2025,lao_canonical_2024,liang_reaching_2026}. We show that we can correct this problem by fine-tuning \ourmodel with a similar functional where 3-body dispersion is included. The OMol-4M subset \cite{levine_open_2025} was recomputed at $\omega$B87M-D4, using the D4 parameters of Friede \textit{et al.}\ \cite{friede_optimally_2023} (this is known as ``wB97M-D4rev'' in ORCA\cite{neese_software_2025} and includes the 3-body ATM correction). Results are shown in Figure \ref{water_density_d4}.

\begin{figure}[ht!]
    \centering
    \includegraphics[width=0.75\textwidth]{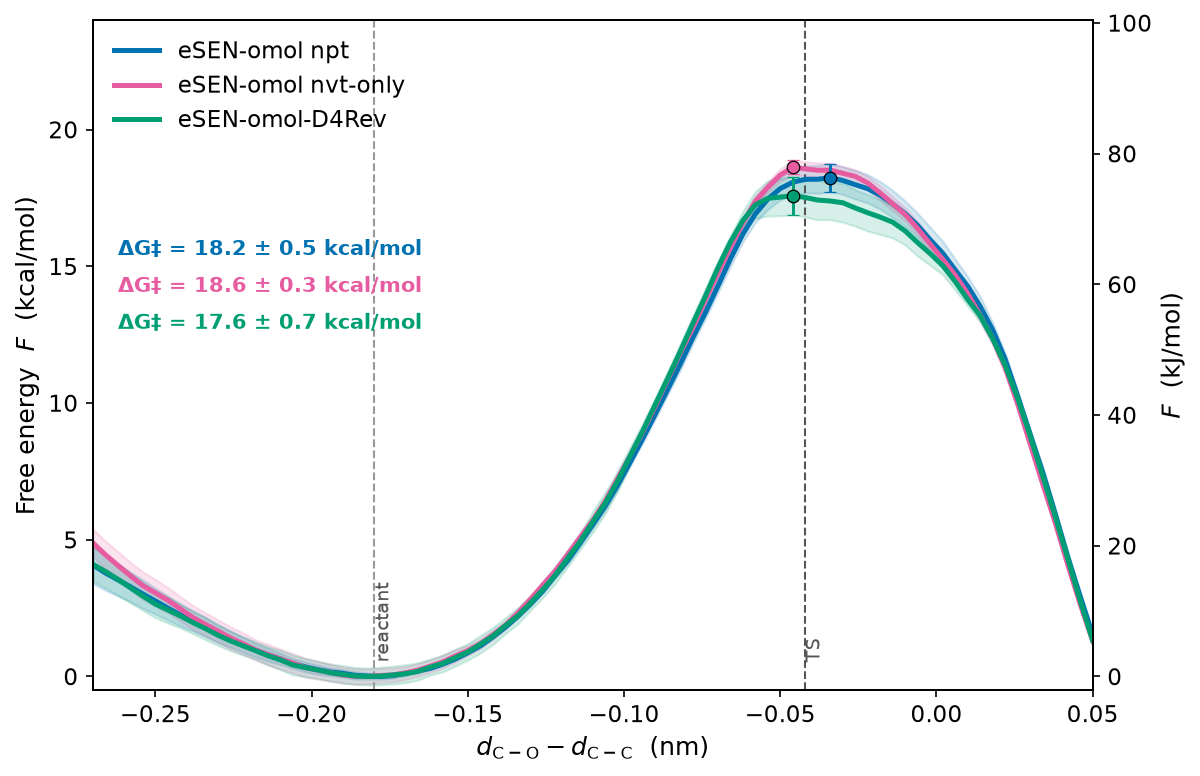}
    \caption{OPES MetaD estimated barrier heights. \textbf{(A)} \ourmodel NPT (conditions used in main text). \textbf{(B)} \ourmodel NVT at 1 g/cm$^3$ without NPT. \textbf{(C)} \ourmodel corrected by finetuning with D4-rev}
    \label{c_mutase_d4}
\end{figure}

In Figure \ref{c_mutase_d4}, we show that using the D4-corrected model, we can closely recover the barrier height compared to the uncorrected model. We also show that we can run all our simulations close to 1 g/cm\textsuperscript{3} densities using NVT alone with no NPT using the uncorrected \ourmodel model to produce nearly identical results. This result is not very surprising: the neglect of 3-body dispersion introduces an extremely systematic shift in the absolute interaction energies of molecules whose effect becomes manifest mainly when these small quantities accumulate at the condensed phase scale; relative energies, such as barrier heights, are expected to be only barely affected.

\clearpage
\newpage
\section{Folding of Polyalanine in Vacuum}\label{app:polyala}
\begin{figure}[ht!]
\centering
\includegraphics[width=1\textwidth]{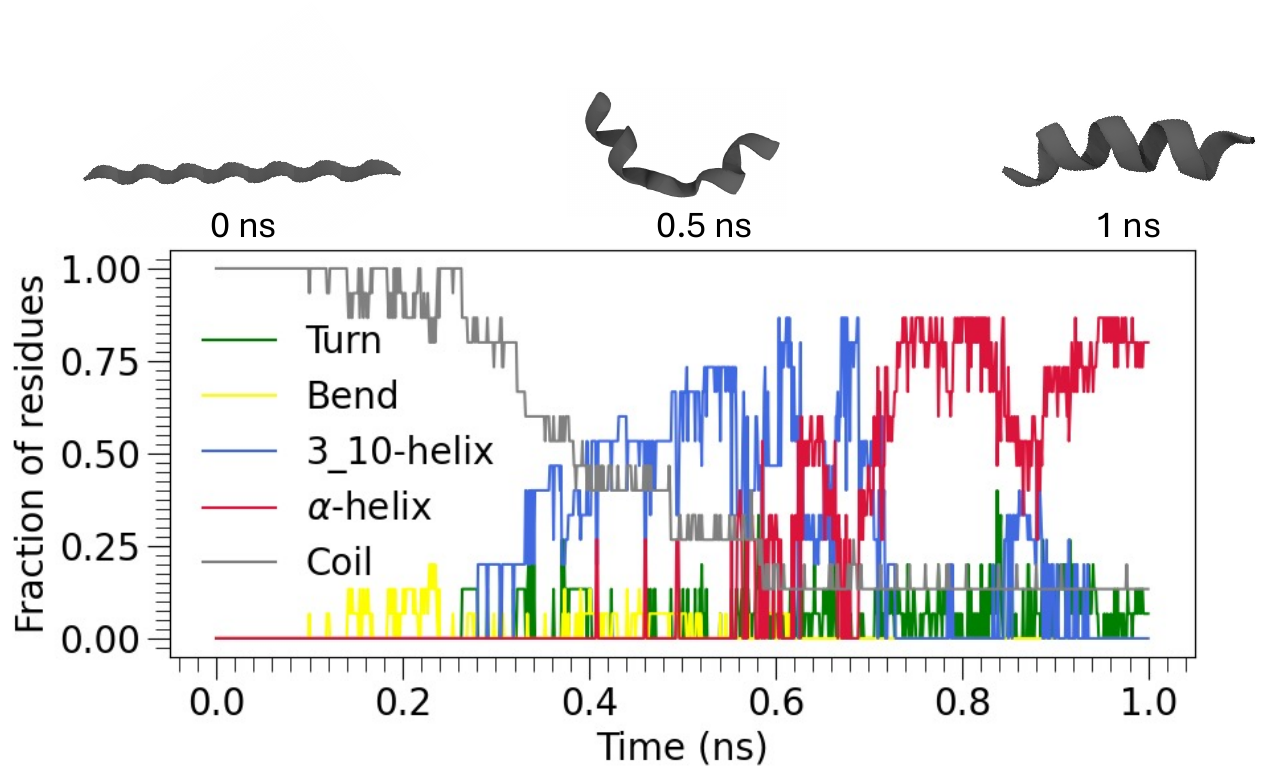}
\caption{ACE-Ala15-NME folds into a helix state from an extended state at 300 K in vacuum. The changes in the fraction of residues contributing to turn, bend, 3$_10$ helix, $\alpha$-helix, and coil secondary structure type during the 1 ns simulation are shown using green, yellow, blue, red, and gray colors, respectively. Snapshots at 0 ns, 0.5 ns, and 1 ns are rendered at the top of the figure using NGLview 3.0.3 \cite{nglview}.}
\label{Fig1}
\end{figure}

The polyalanine peptide with a length of 15 residues (ACE-Ala15-NME) is known to form a stable helix structure at 300 K in vacuum \cite{kovacs2025mace,kabylda2025molecular}. We used this peptide to assess \ourmodel in forming the expected stable helix structure from an extended configuration. We constructed the initial configuration of ACE-Ala15-NME in an extended configuration using AmberTools v24.8 \cite{case2023ambertools}. We then minimized the energy of this structure using the LBFGS algorithm in ASE~\cite{ase-paper}, setting a convergence criterion for the force on all individual atoms to be less than 0.01 eV/\text{\AA} within a maximum of 10,000 steps. Then, we conducted Langevin dynamics for 1 ns with a friction coefficient of 0.01 fs$^{-1}$ and a time step of 1 fs in a vacuum at 300 K using the \ourmodel model s1p1 calculator in ASE~\cite{ase-paper}. The net charge and spin of the system were set to 1 and 0, respectively.  The changes in the secondary structure of ACE-Ala15-NME during the 1 ns simulation, as defined by the Dictionary of Secondary Structure in Proteins (DSSP) \cite{kabsch1983dictionary}, are shown in \Cref{Fig1}. Within 200 ps, the extended coil configuration transitions into a bend configuration and then into a 3$_10$ helix structure. The predominant 3$_10$ helix structure lasts up to 0.7 ns before $\sim$75\% of residues in Ala15 convert to the stable $\alpha$-helix structure. A few of the residues form a turn and retain the coil configuration during the last 0.2 ns of the simulation. The observed variations between the different types of secondary structures with a stable $\alpha$-helix structure at the end of the 1 ns simulation are in good agreement with \cite{kovacs2025mace} and \cite{kabylda2025molecular}.

\end{document}